\RequirePackage{fix-cm}
\documentclass[injp]{article}
\usepackage{graphics}
\usepackage{amsmath}
\usepackage{graphicx}
\usepackage{amssymb}
\usepackage{epsfig}
\usepackage{color}
\usepackage{subfigure}
\usepackage{lineno}
\begin{document}

\parskip 0.2cm
\parindent 1cm

\begin{center}
{\LARGE\bf{Increased metallicity of Carbon nanotubes because of incorporation of extended Stone-Wales' defects: an ab-initio real space approach.}}
\vskip 1cm
{\bf Sujoy Datta{${^\dag}$}},\\
{\baselineskip 7pt Department of Physics, University of Calcutta, 92 Acharya Prafulla Chandra Road, Kolkata 700009, W.B.}\\
\vskip 0.4cm
{\bf Banasree Sadhukhan},\\
{\baselineskip 7pt Department of Physics, Presidency University, 86/1 College Street, Kolkata 700073, W.B.}\\
\vskip 0.4cm
{\bf Chhanda Basu Chaudhuri and Srabani Chakraborty},\\
{\baselineskip 7pt Department of Physics, Lady Brabourne College, P-1/2, Suhrawardy Avenue, Kolkata-700017, India.}\\
\vskip 0.4cm
{\bf Abhijit Mookerjee{${^\ast}$}},\\
{\baselineskip 10pt Department of Condensed Matter and Materials Science, S.N. Bose National  Centre for Basic Sciences, JD-III, Salt Lake, Kolkata 7000098, W.B., India.\\
Visiting Distinguished Professor, Department of Physics, Lady Brabourne College, P-1/2, Suhrawardy Avenue, Kolkata-700017, India.\\
}
\end{center}
 
\vfill \hrule
\vskip 0.1cm
\noindent $^\ast$Corresponding Author (Email: abhijit@bose.res.in; abhijit.mookerjee61@gmail.com)\\  
{\dag} Email: dattasujoy@ymail.com.   
             
\date{Received: date / Accepted: date}

\vskip 0.5cm
\begin{abstract}
 We propose an ab-initio combination of the Linear Muffin-Tin Orbital and the Recursion Methods to study the effect of extended Stone-Wales defects in single layer Carbon nanotubes. We have successfully applied this to zigzag and armchair tubes. The methodology involves no intrinsic mean-field like assumptions or external parameter fitting. As defects proliferate, the low density of states near the Fermi levels of the pristine tubes is filled with defect states. The increase of DOS at the Fermi level leads to enhanced conduction, which indicates enhanced metallicity due to SW defects in the nanotubes. 
\end{abstract}

\baselineskip 12pt
 \section{Introduction}
  From the way graphene is prepared it is quite plausible that disorder is ubiquitous in it. Defect based nano-engineering of graphinic structures have always focused on tubular forms \cite{first}$-$\cite{f3} in addition to sheets. Not only local defects like adatoms or vacancies, but also extended defects like Stone-Wales \cite{1} or patches of other allotropes in a pristine graphene background have been studied \cite{a}$-$\cite{j}. 
  
First-principle studies of extended defects have been beyond the scope of mean-field approaches. The most successful among these, the coherent potential approximation (CPA) can study only very local disorder. The alternative supercell approaches always lead to sharp bands. The singularities of the spectral function are {\sl always} real. This is because on some length scale (size of the super-cell) lattice translation symmetry and Bloch's Theorem still hold. Any estimate of disorder related life-times or band broadening requires smoothening with external parameters. In that sense, the approach is not fully ab-initio. However, the purely real space based recursion method, proposed by the Cambridge group \cite{hhk}, appears to be ideally suited for the study of such extended defects. It is surprising that its use has been rather limited in this particular area of study. We shall present here a recursion based study of such defects and their influence on the electronic structure of Carbon nanotubes.
The recursion method \cite{vol35}$-$\cite{hhk2} introduced by Heine and co-workers is a powerful alternative to the Bloch theorem based techniques and comes to its own when Bloch theorem is violated. A large body of work exist using the recursion method on the study of systems like plane and rough surfaces \cite{mm1}$-$\cite{mm4} and interfaces \cite{vol35}, defects and dopants, disordered alloys \cite{sarma}, amorphous networks \cite{hg,hg2} and frustrated spin systems \cite{amd}.
  \begin{figure*}[]
\centering
\subfigure[]{\framebox{\includegraphics[width=3cm,height=3cm]{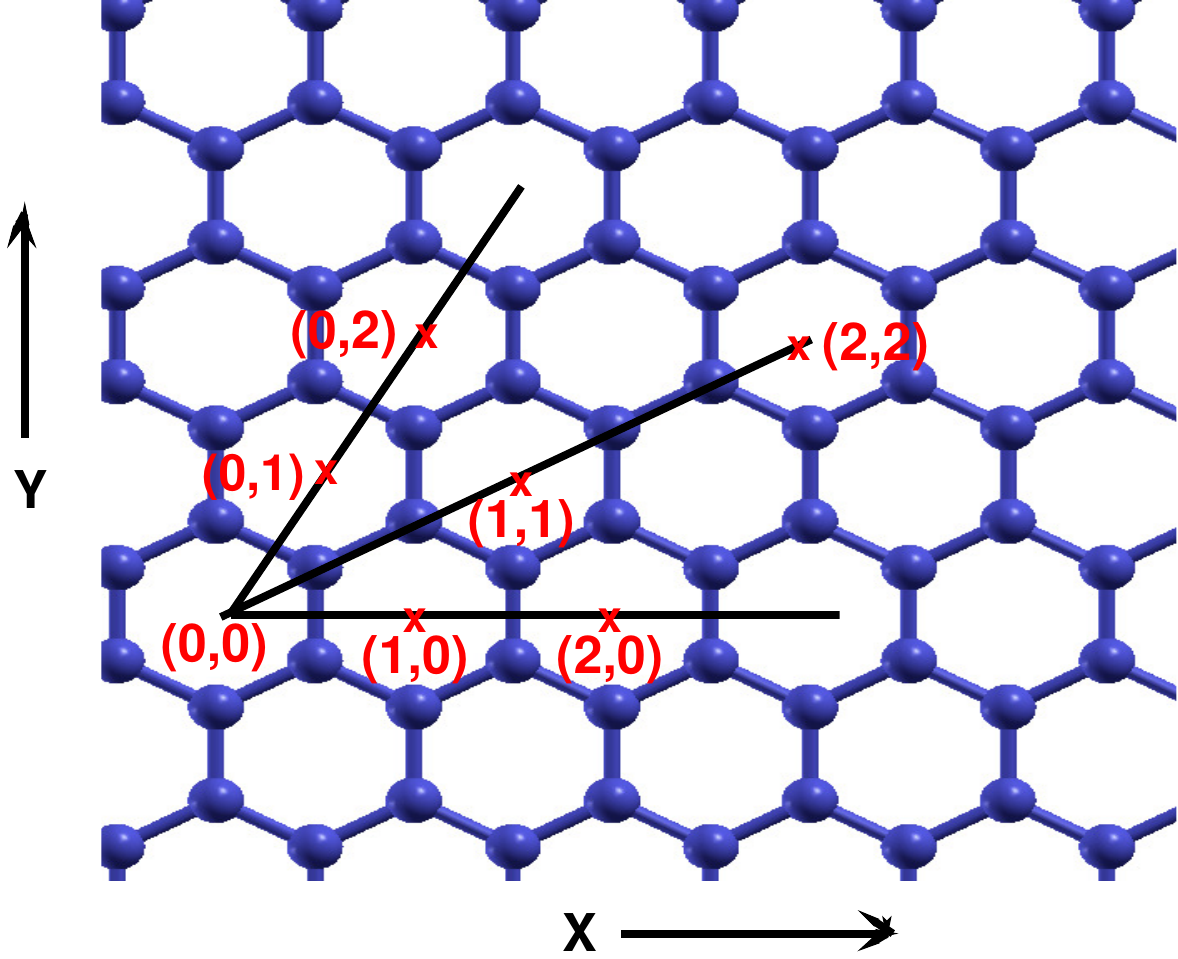}}}
\subfigure[]{\framebox{\includegraphics[width=3cm,height=3cm]{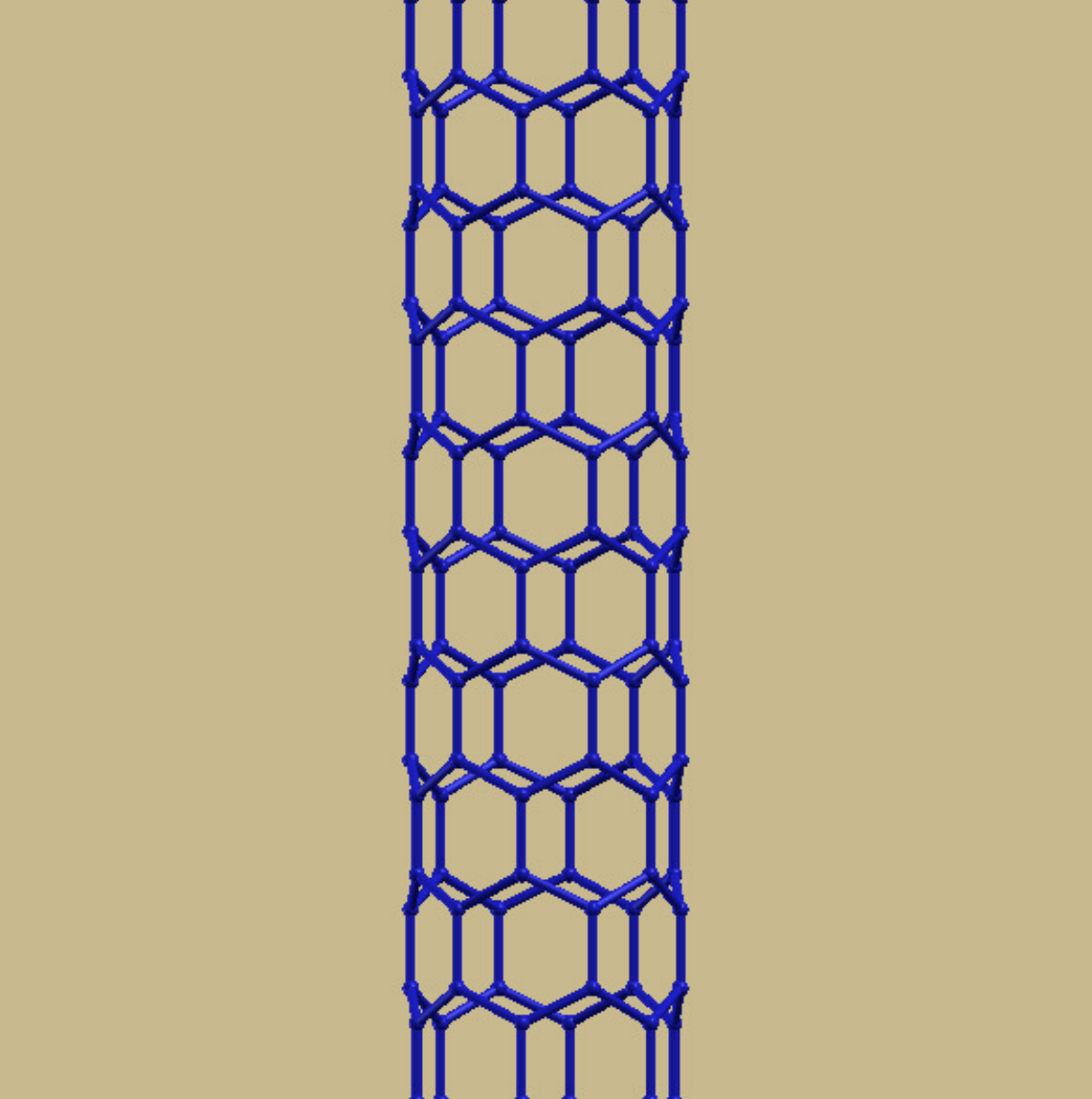}}}
\subfigure[]{\framebox{\includegraphics[width=3cm,height=3cm]{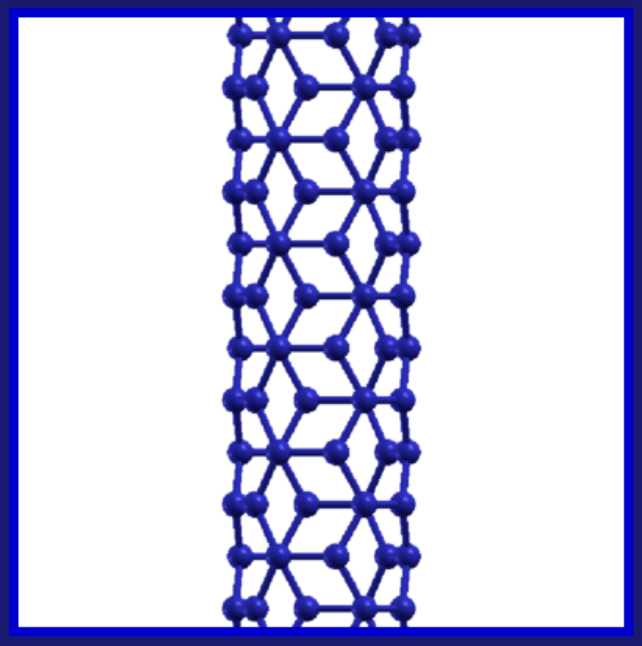}}}
\subfigure[]{\framebox{\includegraphics[width=3cm,height=3cm]{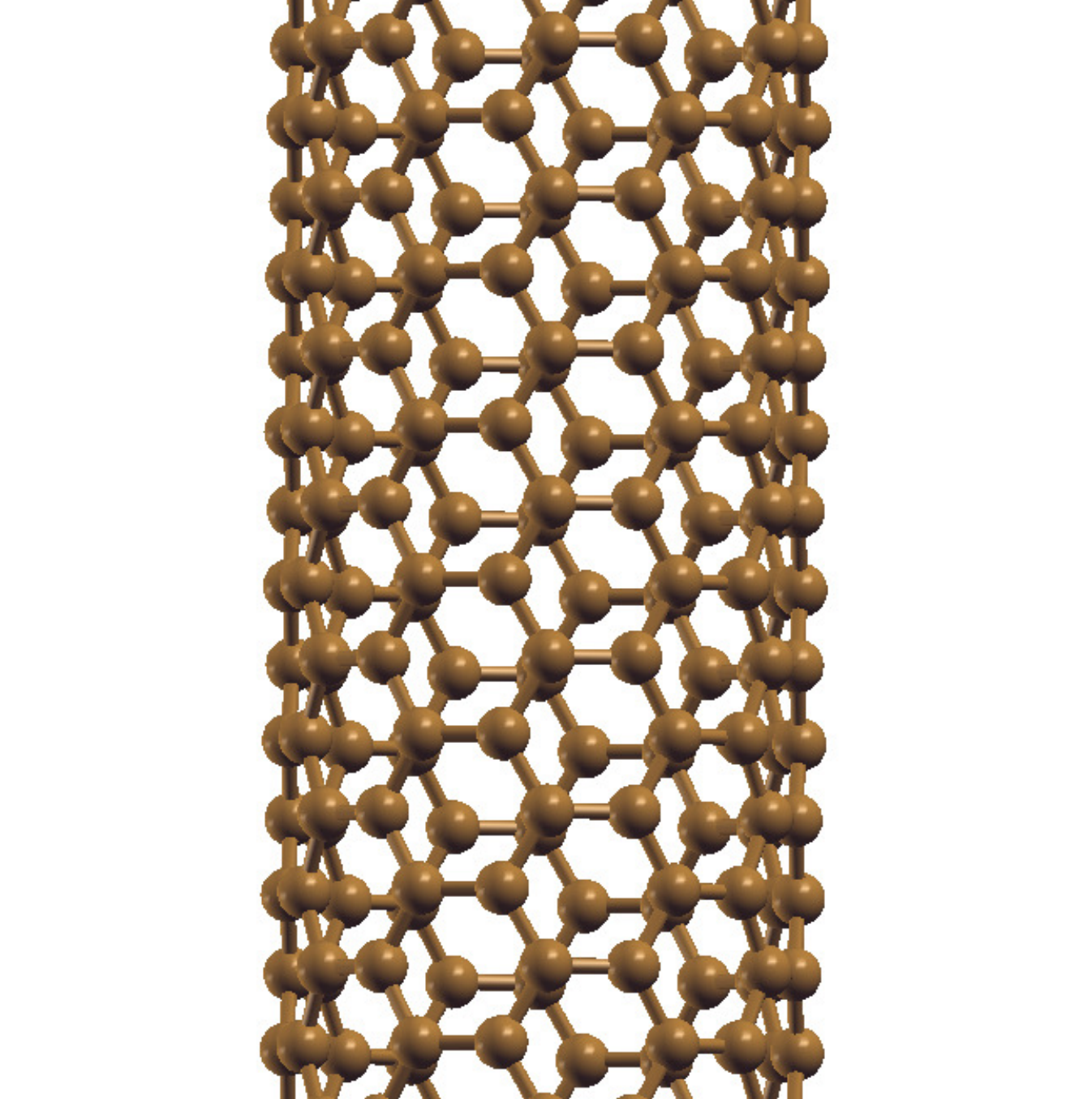}}}
\vskip -0.4cm
\caption{\baselineskip 8pt
\small\rm  (a) An ordered graphene lattice; (b) a (3,0) zigzag nanotube of radius 4.92${\rm \AA}$; (c) a (3,3) armchair nanotube of radius 7.57${\rm \AA}$;  (d) a (6,6) armchair nanotube of radius 8.14${\rm \AA}$.
\label{grph}}
\end{figure*}

  \subsection{Nanotube and Stone-Wales Defects}
  The Fig.\ref{grph}(a) shows a pristine two-dimensional honeycomb structure. If we fold this about y-axis then we obtain a zigzag type nanotube (Fig.\ref{grph}(b)). Whereas folding about x-axis leads to a nanotube of the armchair type \cite{charlier}$,$ \cite{20}$-$\cite{27} (Fig.\ref{grph}(c)). A general identification of the type of nanotube is given by chiral vectors (0,n) and (n,0), not the standard orthogonal x,y-axes. A zigzag type nanotube is described by $(n,0)$ and armchair type by $(n,n)$. Fig.\ref{grph}(b) is a (6,0) nanotube of radius 4.92${\rm \AA}$, Fig.\ref{grph}(c) is a (3,3) nanotube of radius 7.57${\rm \AA}$ and Fig.\ref{grph}(d) is a (6,6) nanotube of radius 8.14${\rm \AA}$ . Electronic properties of these two types have been investigated by other techniques before and accordingly zigzag type nanotubes are called  semiconductor nanotubes and armchair types are known as metallic nanotubes \cite{20,22,29}. Only the orientation or, generally speaking, the geometry decides the metallic or semiconducting behaviour of nanotubes. This made us interested in geometrical disorders in these. We studied the electronic properties of both types with and without Stone-Wales (S-W) defects.

  \begin{figure}[b!]
\centering
\subfigure[]{\framebox{\includegraphics[width=3.cm,height=4cm]{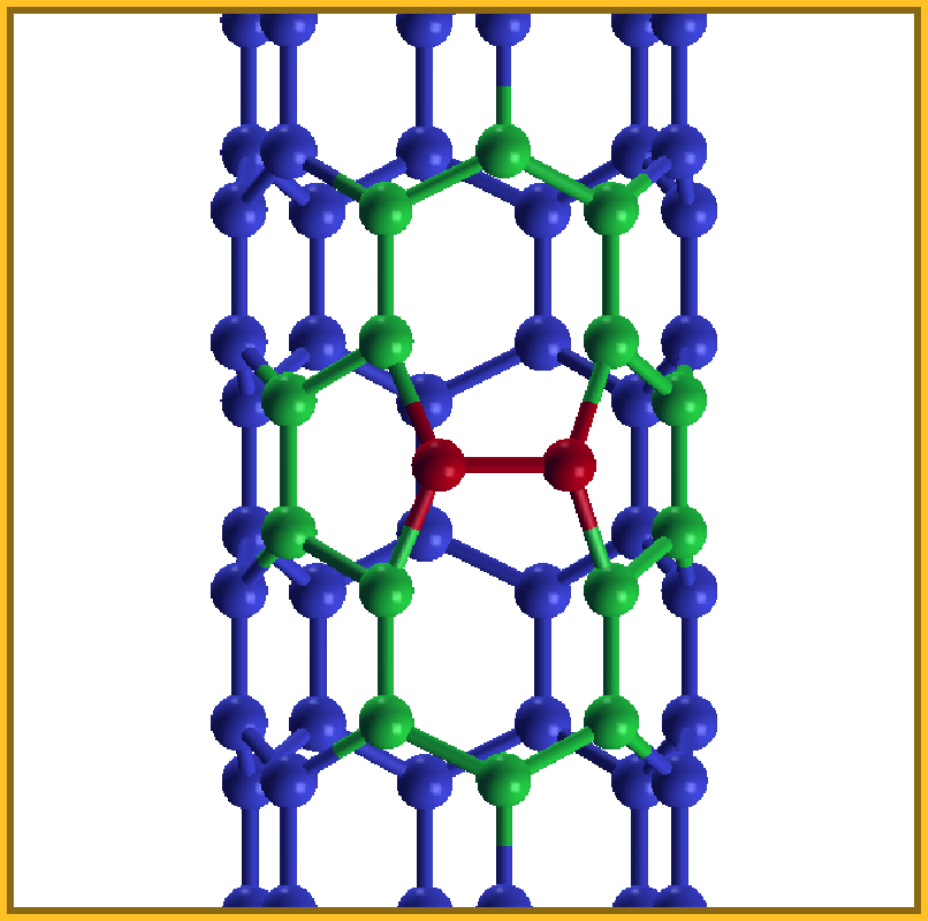}}}
\subfigure[]{\framebox{\includegraphics[width=3.cm,height=4cm]{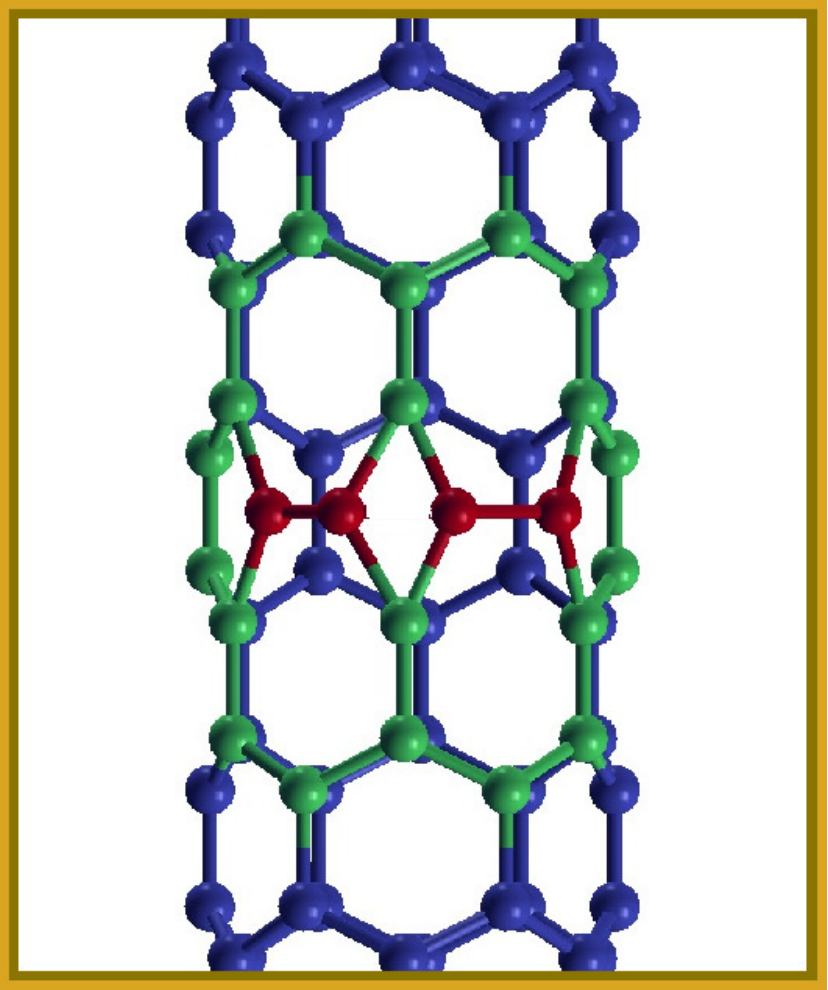}}}
\caption {\baselineskip=8pt \small\rm (a) Single and (b) Double Stone-Wales defects in a zigzag type nanotube.\label{defects}}
\end{figure}

   Stone-Wales (S-W) defects \cite{2}$-$\cite{17} are  purely geometrical extended defects which occur due to rotation of two neighbouring atoms shown in red in Fig.\ref{defects}, followed by rebonding of the dangling bonds so produced. A single S-W defect spreads over sixteen atoms and transforms four adjacent hexagonal cells into two pentagonal and two heptagonal cells are shown in green in Fig.\ref{defects}(a). In a double S-W defect, twenty-four atoms are involved and six adjacent hexagonal cells transform into one rhombic, two pentagonal and four heptagonal cells are shown in green in Fig.\ref{defects}(b). 


  \section{Methodology}
  The elements of the representation of the Tight-Binding Anderson Hamiltonian \cite{and} have been obtained first from an ab-initio, self-consistent, Tight-Binding Linear Muffin-tin Orbitals technique (TB-LMTO) \cite{lmto2}$-$\cite{lmto4}. The resulting Hamiltonian basis is multi-orbital with $s$,$p$ and $d$ orbitals. Our interest is in the $p_z$ sector alone. The N-th order muffin-tin orbital (NMTO method) \cite{nmto,nmto2} and massive downfolding are used to obtain the $p_z$ Hamiltonian.
  Geometrically we find the angle between the plane containing the three nearest neighbours and any of C-C bond is very small (Maximum is ${9.7^0} $ for (3,3) nanotube.). So, the tube is considered to be locally flat and the calculated values of Hamiltonian parameters for the flat graphene sheet were used locally.

  \begin{table*}[htp]\label{ht}
\caption{\bf {The NMTO Hamiltonian for the $p_z$ sector of C}}
\begin{center}
\begin{tabular}{|c|c|}
\hline
$\varepsilon$ (eV) & t (eV) \\ \hline
-0.291 & -2.544\\ \hline
\end{tabular}
\end{center}
\end{table*}

  \begin{equation}
 \widehat{H} = \sum_{i} \varepsilon_i |i\rangle\langle i| + \sum_{\left\lbrace i,j\right\rbrace } t_{ij} 
 \left\{ |i\rangle\langle j| + |j\rangle\langle i|\right\}
\end{equation}
where the site labelled basis $\{i\}$ is used for representation, $\{i,j\}$ refer to nearest neighbour pairs in the material and $\langle i| j \rangle = \delta_{ij}$ .

The recursion method transforms the set of basis from $\{|i\rangle\}$ to $\{|u_n\rangle\}$, which converts the tight-binding Hamiltonian into a Jacobian matrix. It has been described in great detail by Haydock {\sl et.al.} \cite{vol35}$-$\cite{hhk2}.
Note that there is absolutely no symmetry restriction on the matrix: no translation symmetry or local geometric symmetries. It is enough that the basis is denumerable. The recursion method does not necessarily invoke the Bloch Theorem and therein lies its strength.  
\begin{eqnarray}
 |u_1\rangle & = & |k\rangle\quad \mbox{(our choice from the set ${\{|i\rangle\}}$)}\nonumber\\
 |u_{n+1}\rangle & = & \widehat{H} |u_{n}\rangle - \alpha_{n} |u_{n}\rangle - \beta^2_{n}|u_{n-1}\rangle\\
 \phantom{x} & & \phantom{x}\nonumber \\
 \noindent {where,}\nonumber \\
   \alpha_n & =&  \langle u_n|\widehat{H}|u_n\rangle/ \langle u_n|u_n\rangle\nonumber\\
   \beta^2_{n+1} & =& \langle u_{n+1}|u_{n+1}\rangle/\langle u_n|u_n\rangle
\end{eqnarray}

 The diagonal element of the Green's function in matrix representation is:
 \begin{eqnarray}
   G_{nn}(z)& =& \langle u_n|\left(z\widehat{I}-\widehat{H}\right)^{-1}| 
 u_n\rangle  \nonumber\\
 & = &\frac{1}{\displaystyle z-\alpha_1- \frac{\beta_1^2}{\displaystyle z-\alpha_2-\frac{\beta_2^2}{\displaystyle \ddots z-\alpha_N-T(z)}}} \nonumber\\  
\end{eqnarray} 

The terminator $T(z)$ is built up from the asymptotic behaviour of initial coefficients together with the known singular points in the spectrum of Green function.
 
\begin{figure*}[t!]
\centering
\vskip 1cm
\subfigure[]{\framebox{\includegraphics[width=3cm,height=3.0cm]{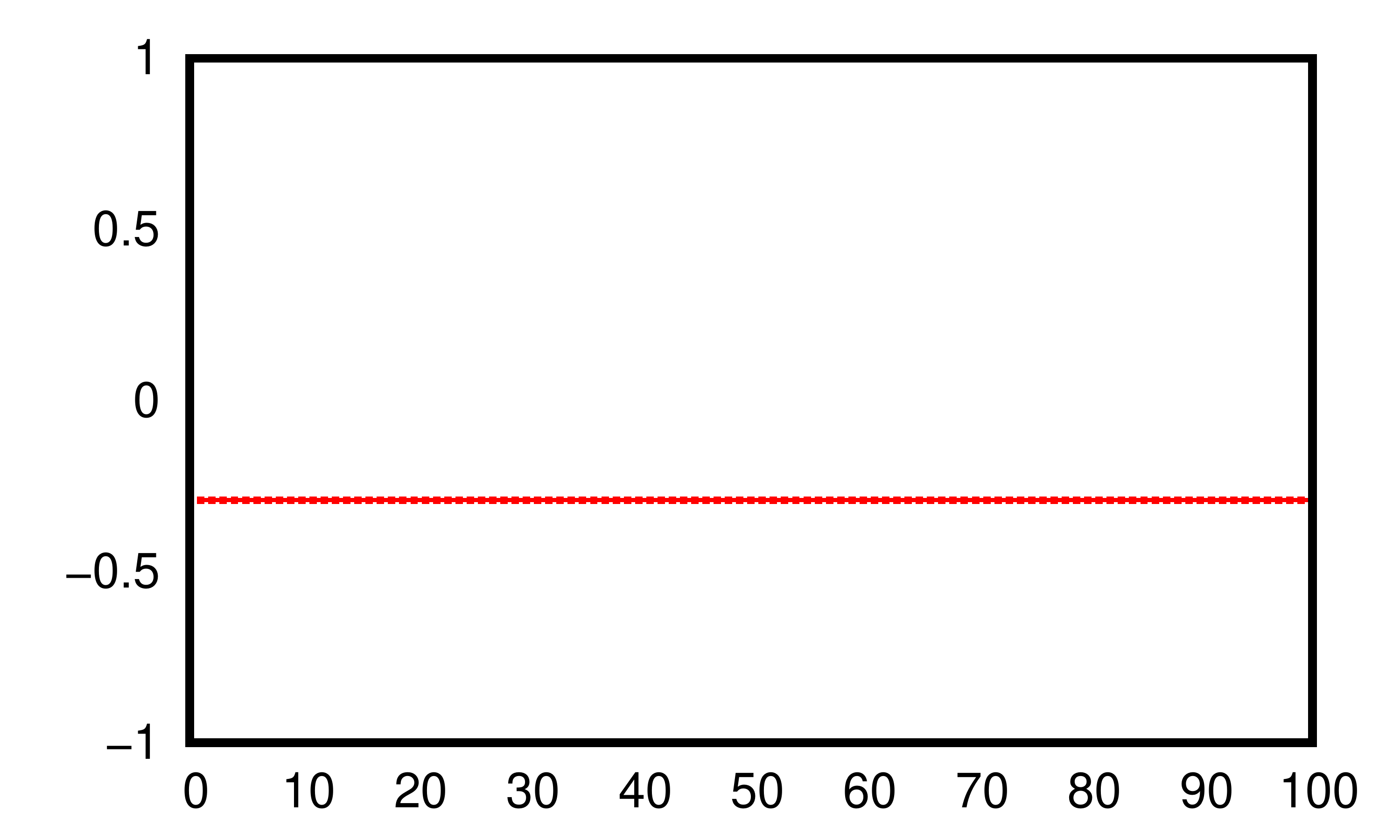}}}
\subfigure[]{\framebox{\includegraphics[width=3cm,height=3.0cm]{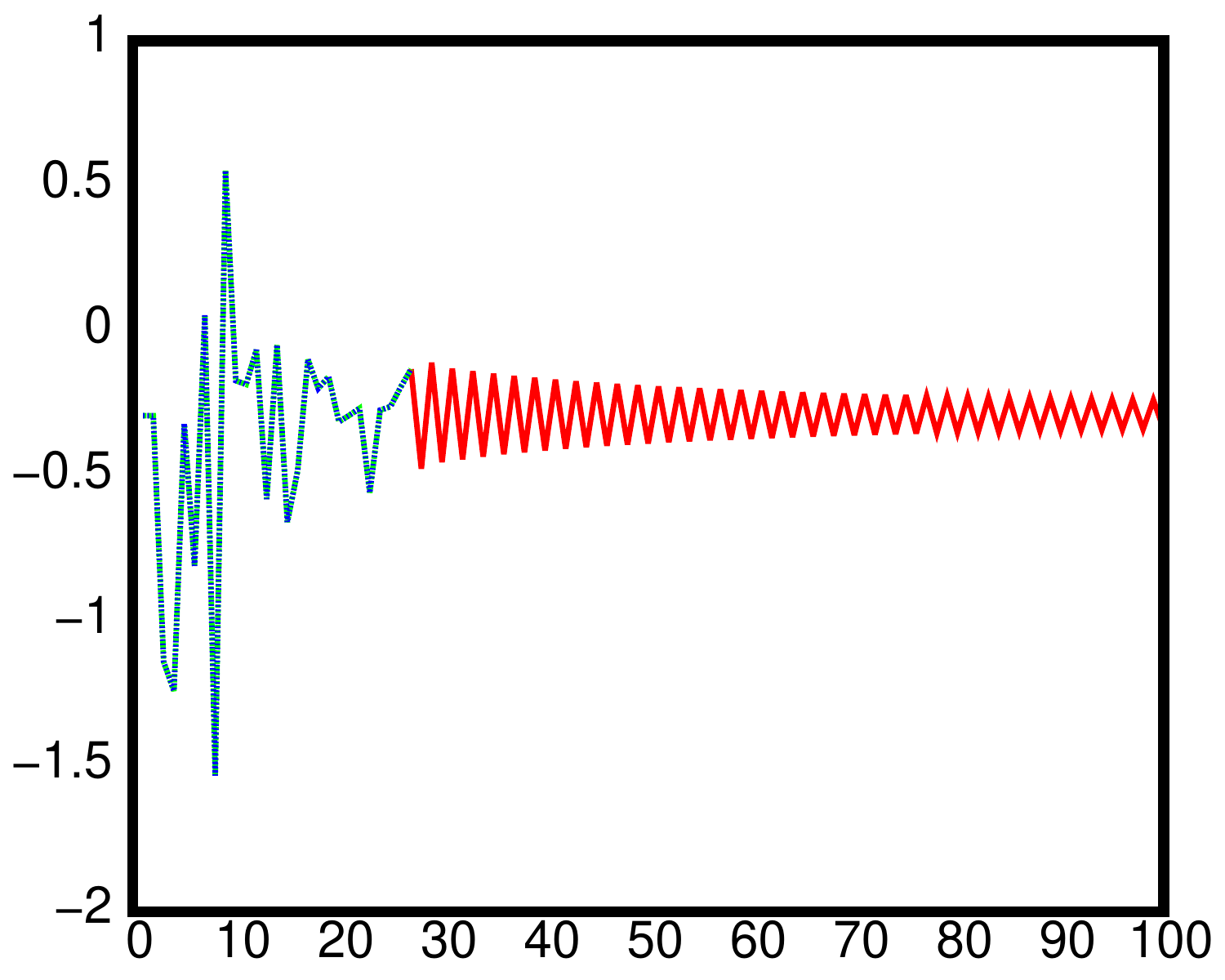}}}
\subfigure[]{\framebox{\includegraphics[width=3cm,height=3.0cm]{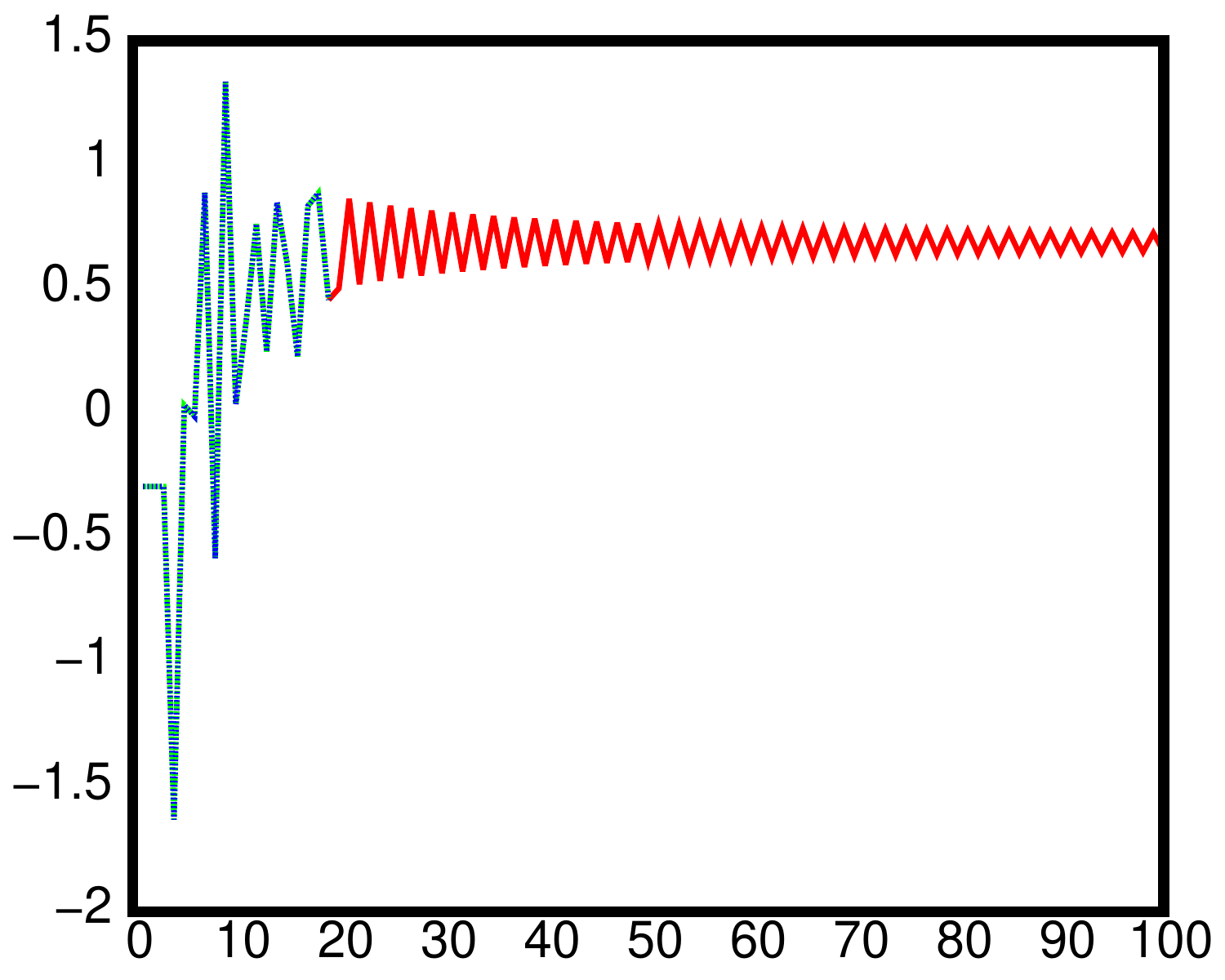}}}\\
\subfigure[]{\framebox{\includegraphics[width=3cm,height=3.0cm]{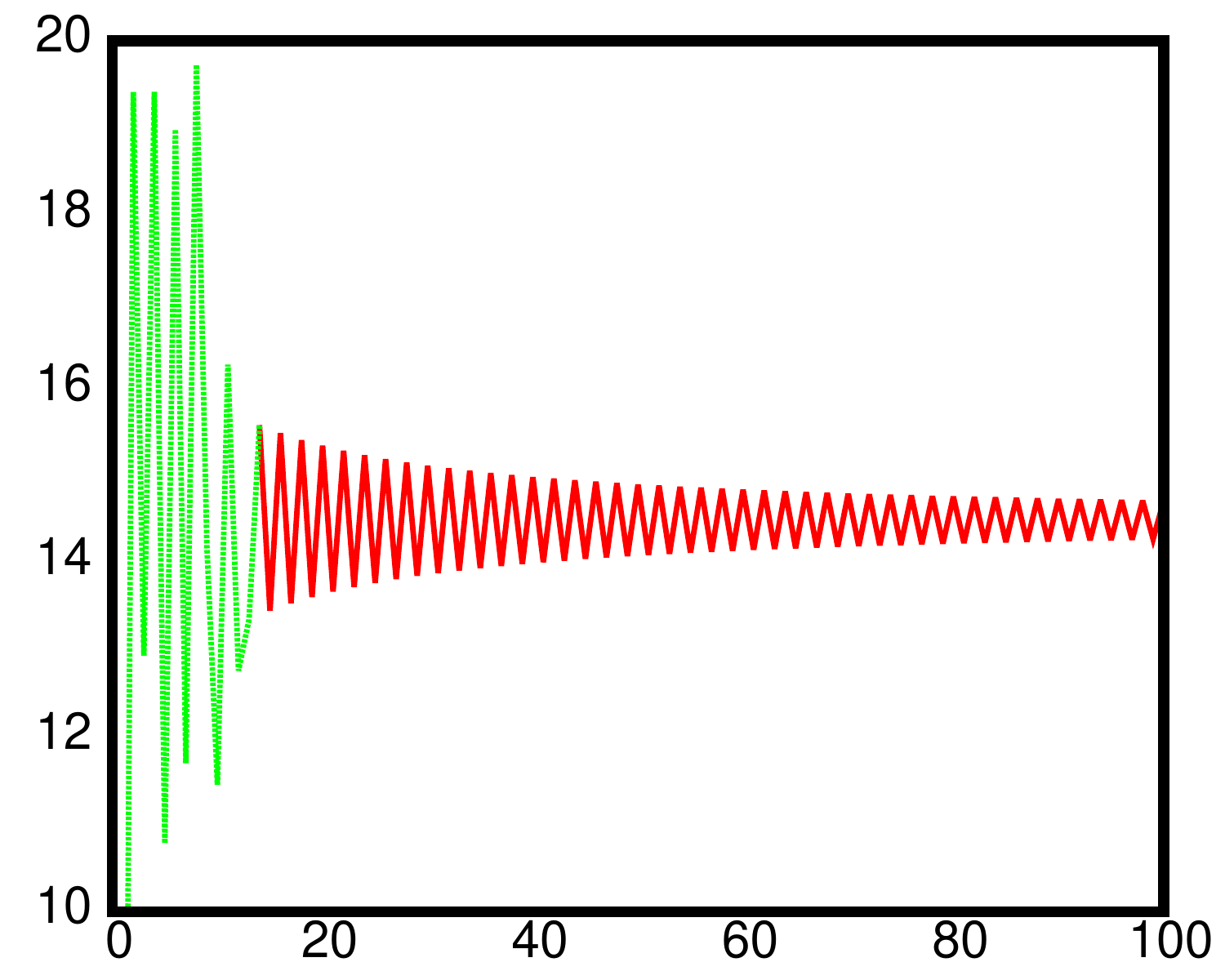}}}
\subfigure[]{\framebox{\includegraphics[width=3cm,height=3.0cm]{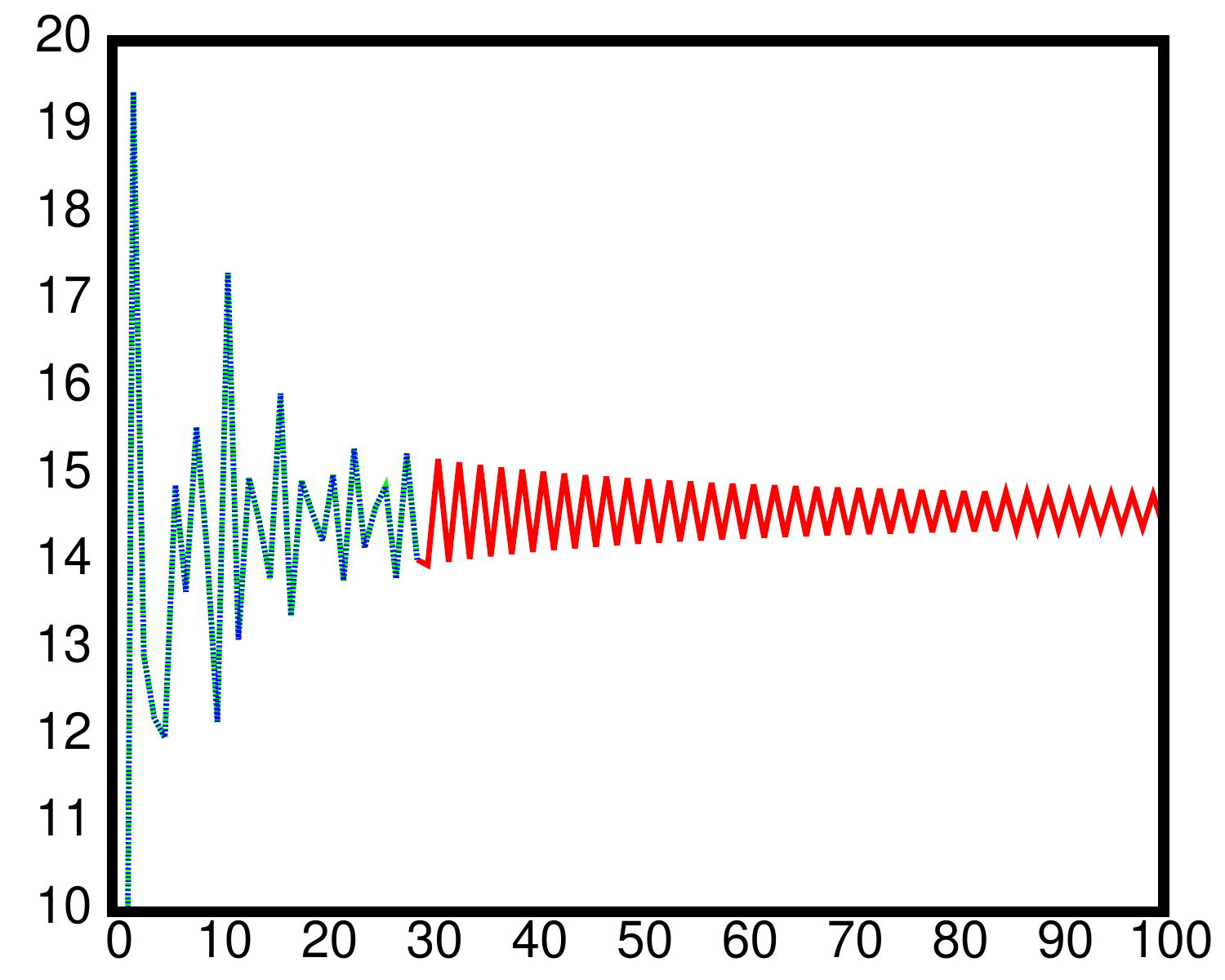}}}
\subfigure[]{\framebox{\includegraphics[width=3cm,height=3.0cm]{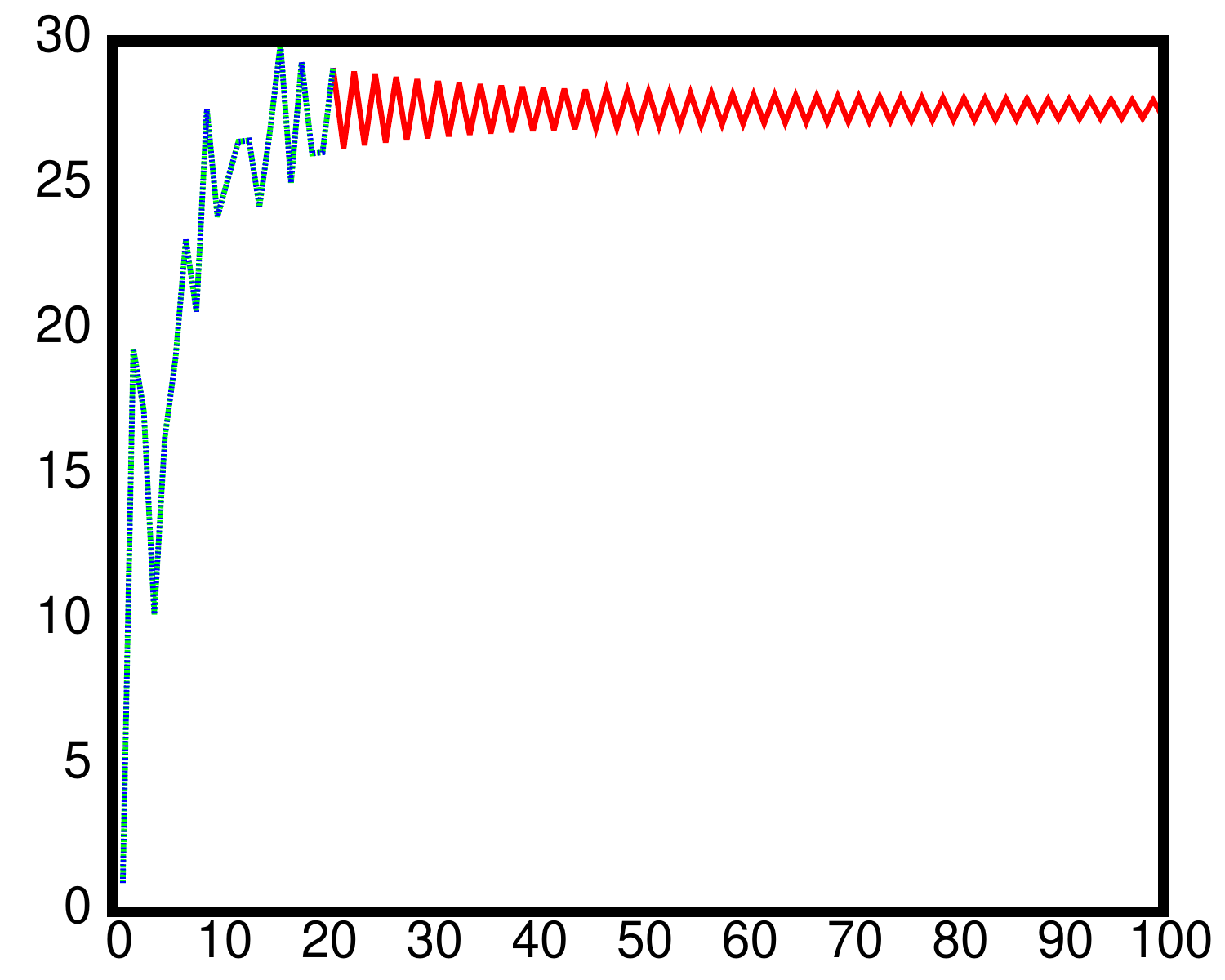}}}
\caption{\label{ab} $\alpha$'s for Zigzag Nanotubes:(a) pristine; (b) with single S-W defect; (c) with double S-W defect.
$\beta$'s for Zigzag Nanotubes:(d) pristine; (e) with single S-W defect; (f) with double S-W defect.}
\end{figure*}

The local density of states (LDOS) and total density of states (TDOS) are defined as 
\begin{eqnarray}
n_0(E) & =&  -(1/\pi) \lim_{\delta\rightarrow 0} {\rm Im}  \ G_{00}(z)\nonumber\\
n(E) & = & -(1/\pi) \lim_{\delta\rightarrow 0} {\rm Im} \ {\rm Tr}\ G(z)
\end{eqnarray}

 Numerically, we can not iterate the continued fraction for infinite number of steps, so, we must terminate this continued fraction at some stage. A good description of terminators is given by Muller and Viswanath \cite{mb}. The manner in which the coefficients behave as $n\rightarrow\infty$ reflects upon the bandwidths and integrable singularities on the compact spectrum
 
 We have used terminators of the type:
\begin{equation}
T(z)=\frac{2\pi}{ {E_0}^{4}} \vert E \vert ({E_0}^2-E^2)
\end{equation} 
with expansion coefficients
\[\beta_{2n}^2 = \frac{E_0^2n}{2(2n+1)}\quad {\rm and} \quad \beta_{2n+1}^2=\frac{E_0^2(n+2)}{2(2n+3)}\]
where $\pm E_0$ are the band edges.
Using terminators the $\{\alpha_n\}$ and $\{\beta_n\}$ are plotted against $n$ in Fig.\ref{ab} for some cases. 

 \section{Results and Discussions}

  This section is divided into two parts: the first presents a discussion on how the calculated values of the Hamiltonian parameters $\varepsilon$ and $t$ obtained from TB-LMTO/NMTO reproduce results in the case of nanotubes with and without defects. In the second we present a comparative discussion between the LDOS for zigzag type and armchair type nanotubes with S-W defects.  
  
 In earlier works the energy was usually expressed in terms of $(E-\epsilon)/E_b$ where $E_b=2Zt$ is the bandwidth and $Z$ is the number of nearest neighbours of each atom. So, they chose $\epsilon=0$ and $t=1$. However, when defects are introduced $\alpha_n$ varies with $n$ and this shift of the origin to a constant value of $\epsilon$ is not possible any more. However the scaling of energy by the bandwidth ($2Zt$) still remains valid. Also in such cases, the symmetry of the DOS is lost because the odd moments of LDOS are functions of $\{\alpha_n\}$ \cite{vol35}. In our study, the Hamiltonian parameters $\epsilon$ and $t$ are obtained from an initial TB-LMTO followed by a massive down-folding to obtain their effective $p_z$ parts \cite{am1}. 
 \begin{figure*}[]
\centering
\vskip -3cm
\subfigure[]{\framebox{\includegraphics[width=2.5cm,height=3cm]{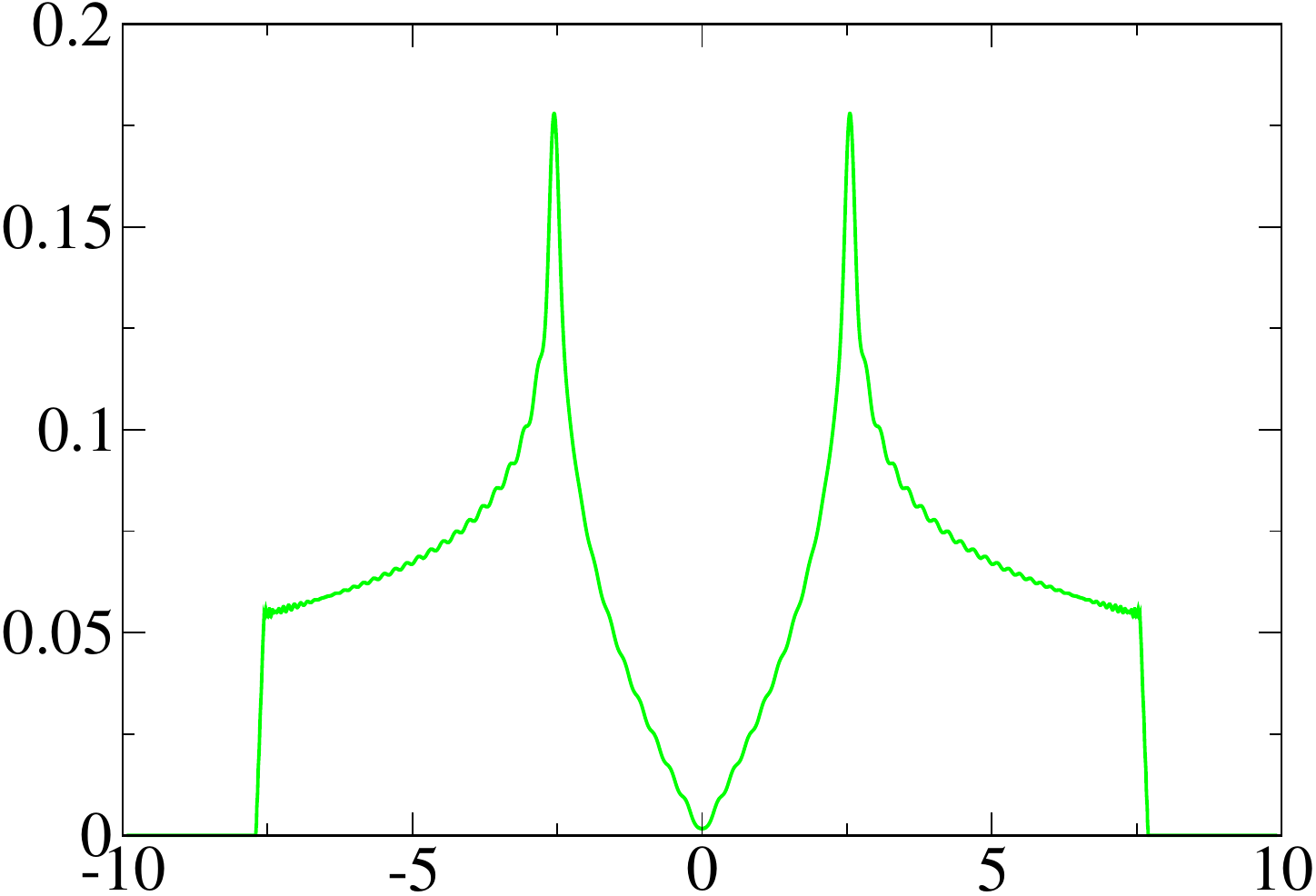}}}
\subfigure[]{\framebox{\includegraphics[width=2.5cm,height=3cm]{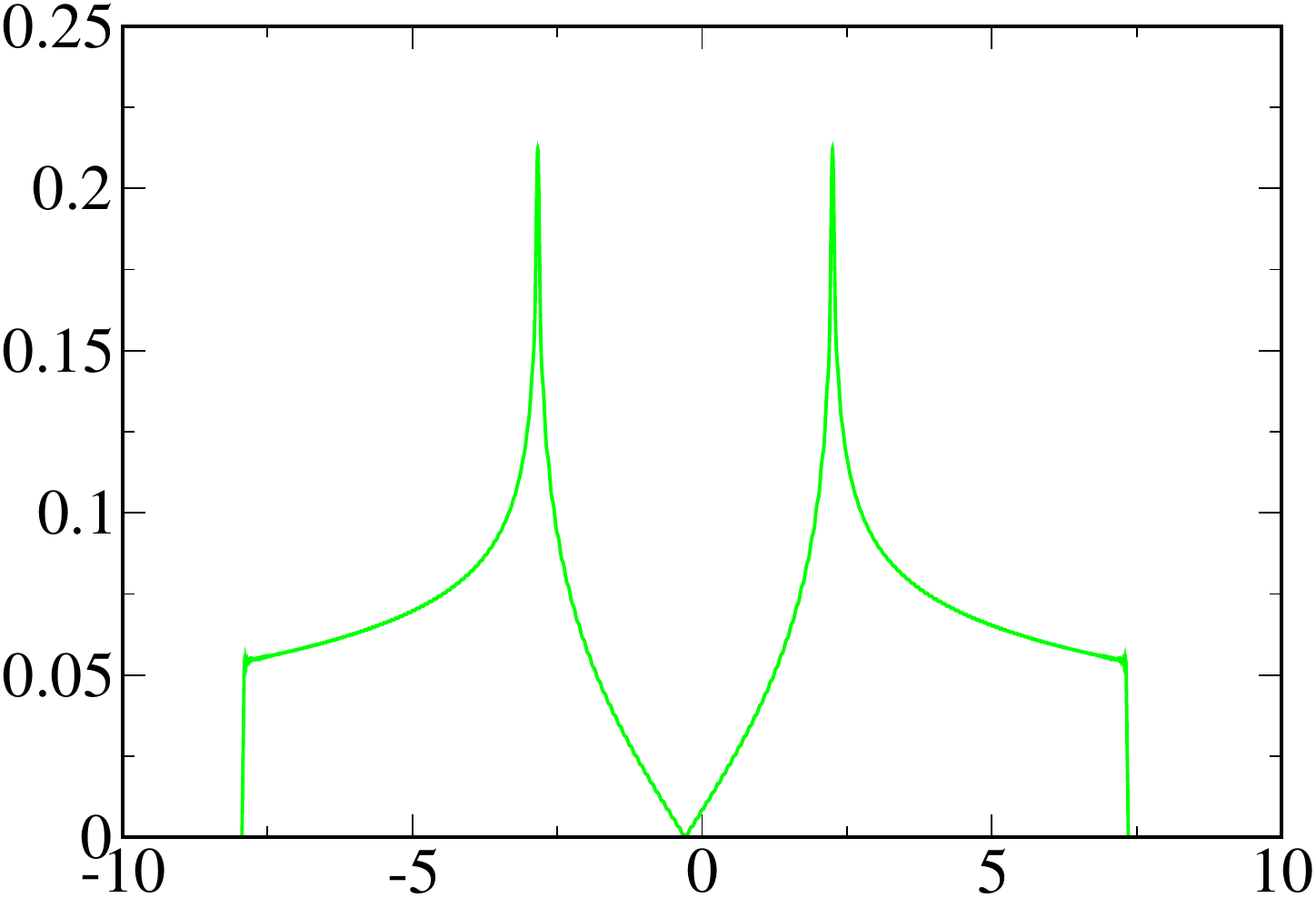}}}
\subfigure[]{\framebox{\includegraphics[width=2.5cm,height=3cm]{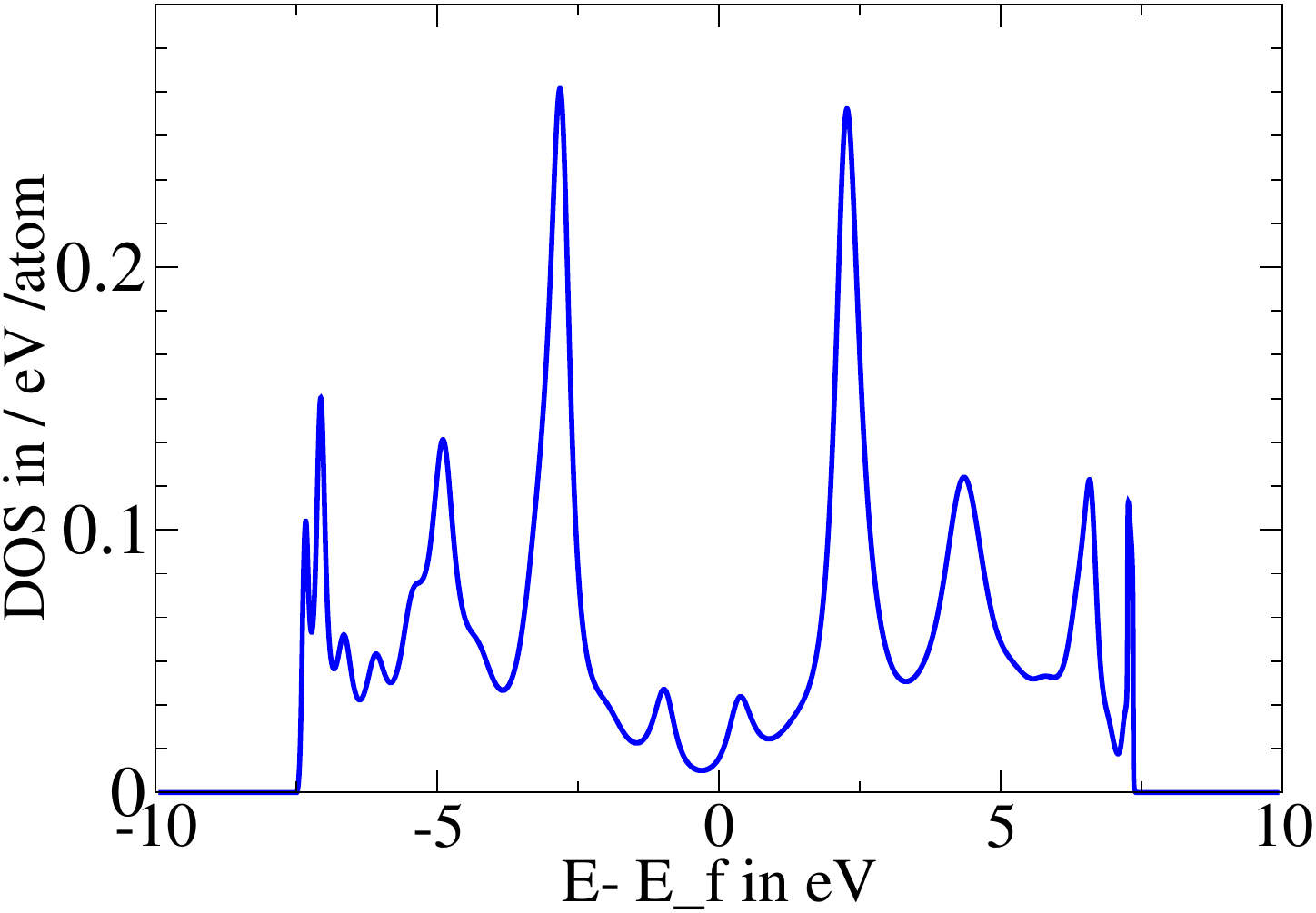}}}
\subfigure[]{\framebox{\includegraphics[width=2.5cm,height=3cm]{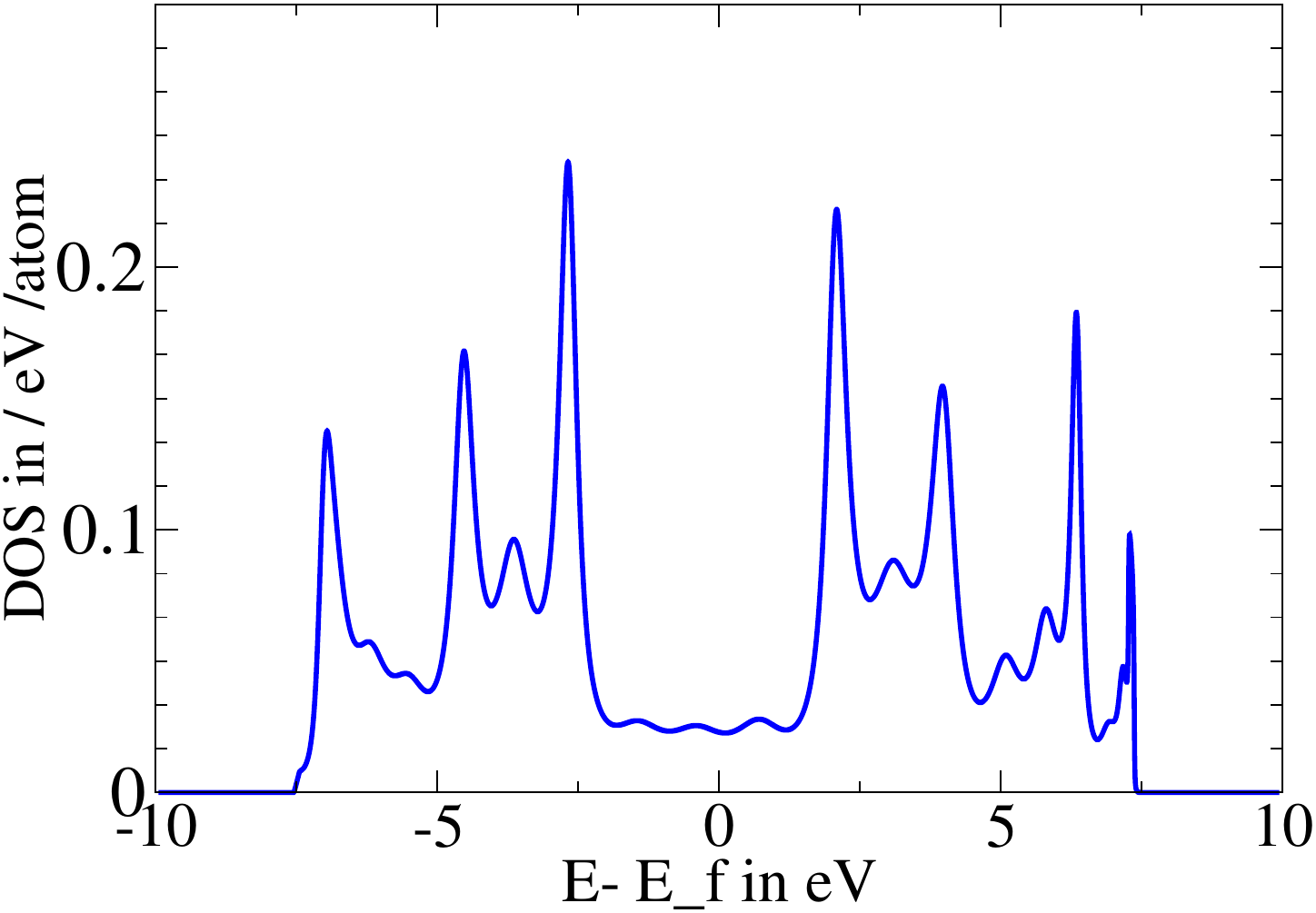}}}
\subfigure[]{\framebox{\includegraphics[width=2.5cm,height=3cm]{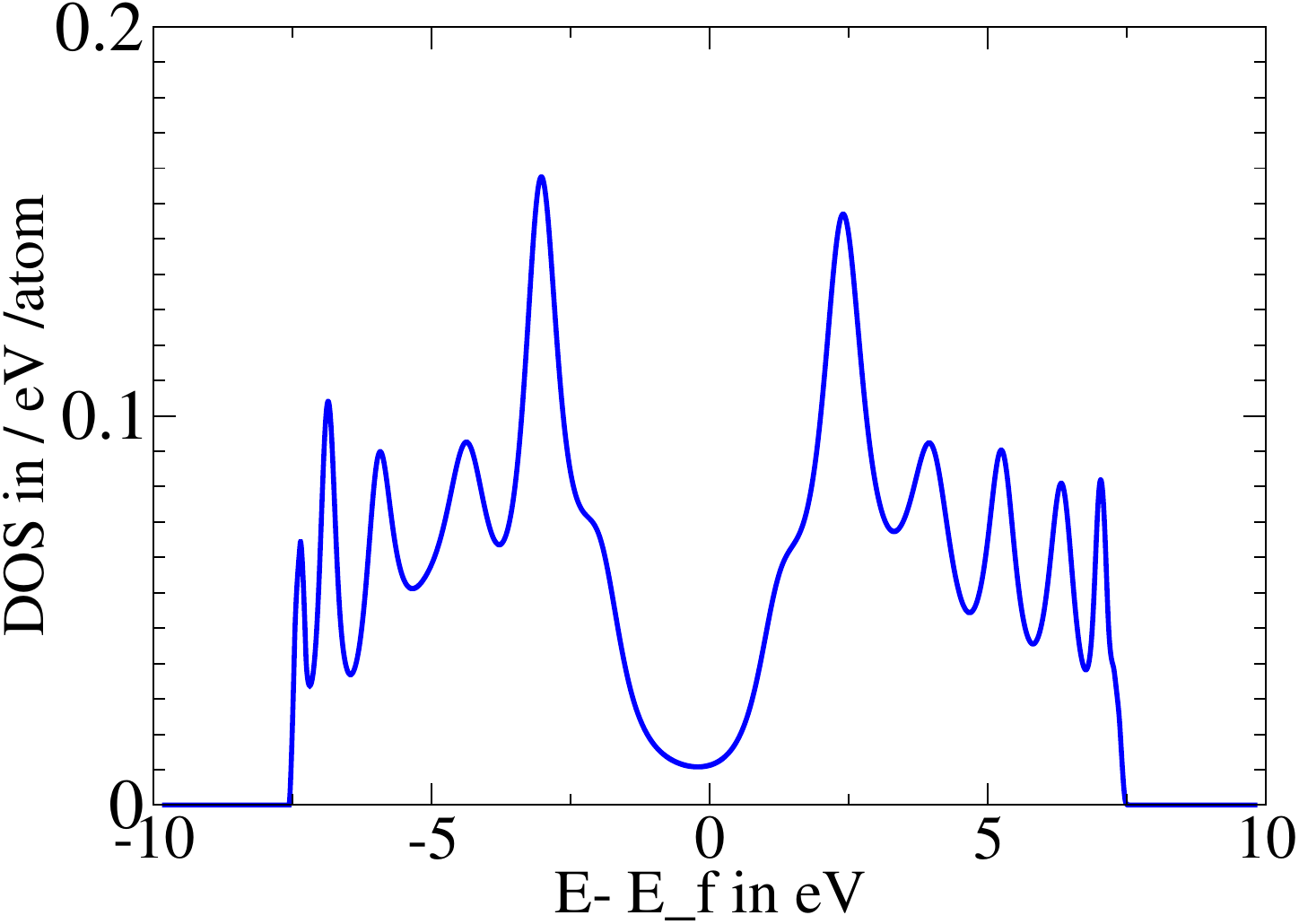}}}\\
\subfigure[]{\includegraphics[width=3.7cm,height=4cm]{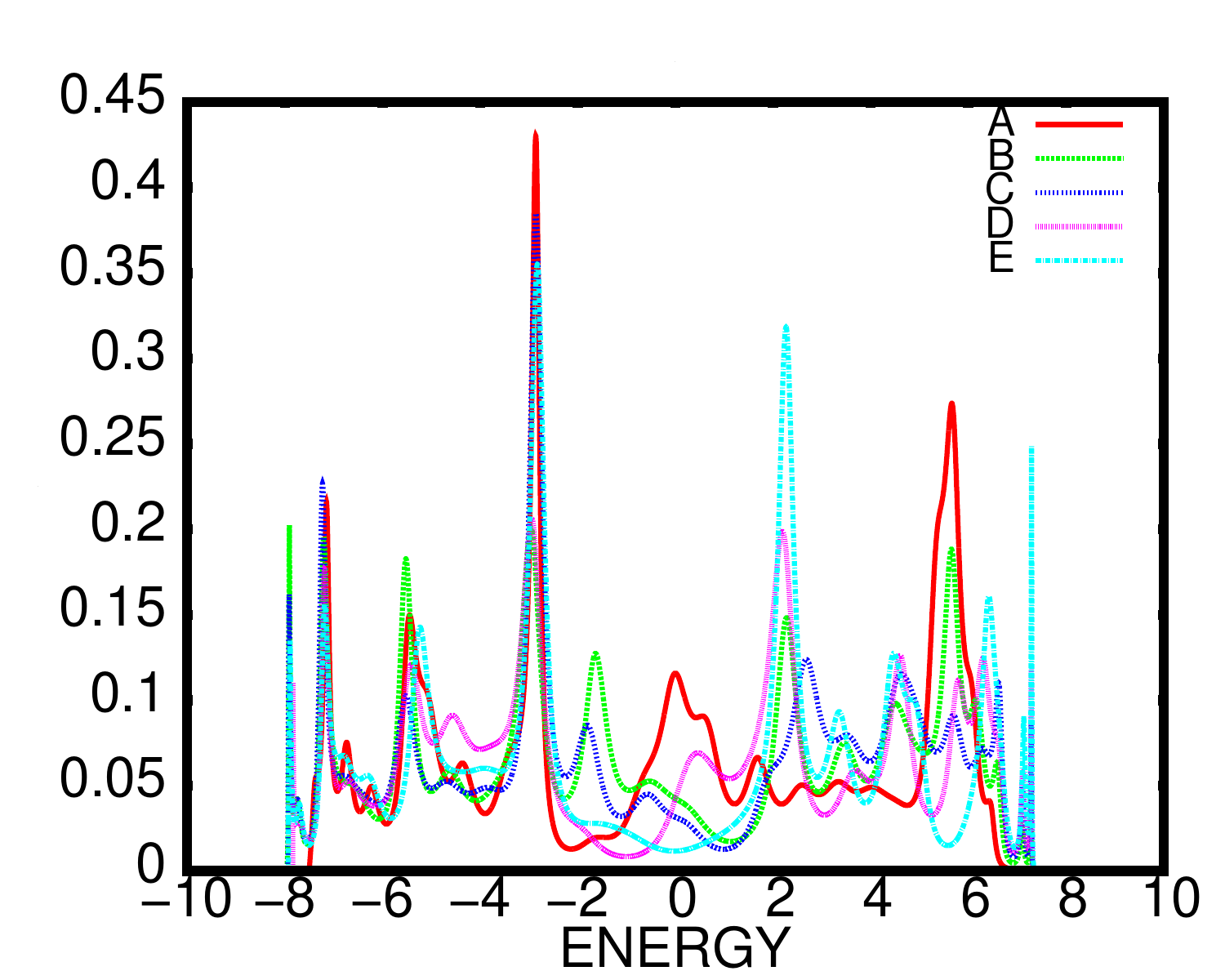}}
\subfigure[]{\includegraphics[width=3.7cm,height=4cm]{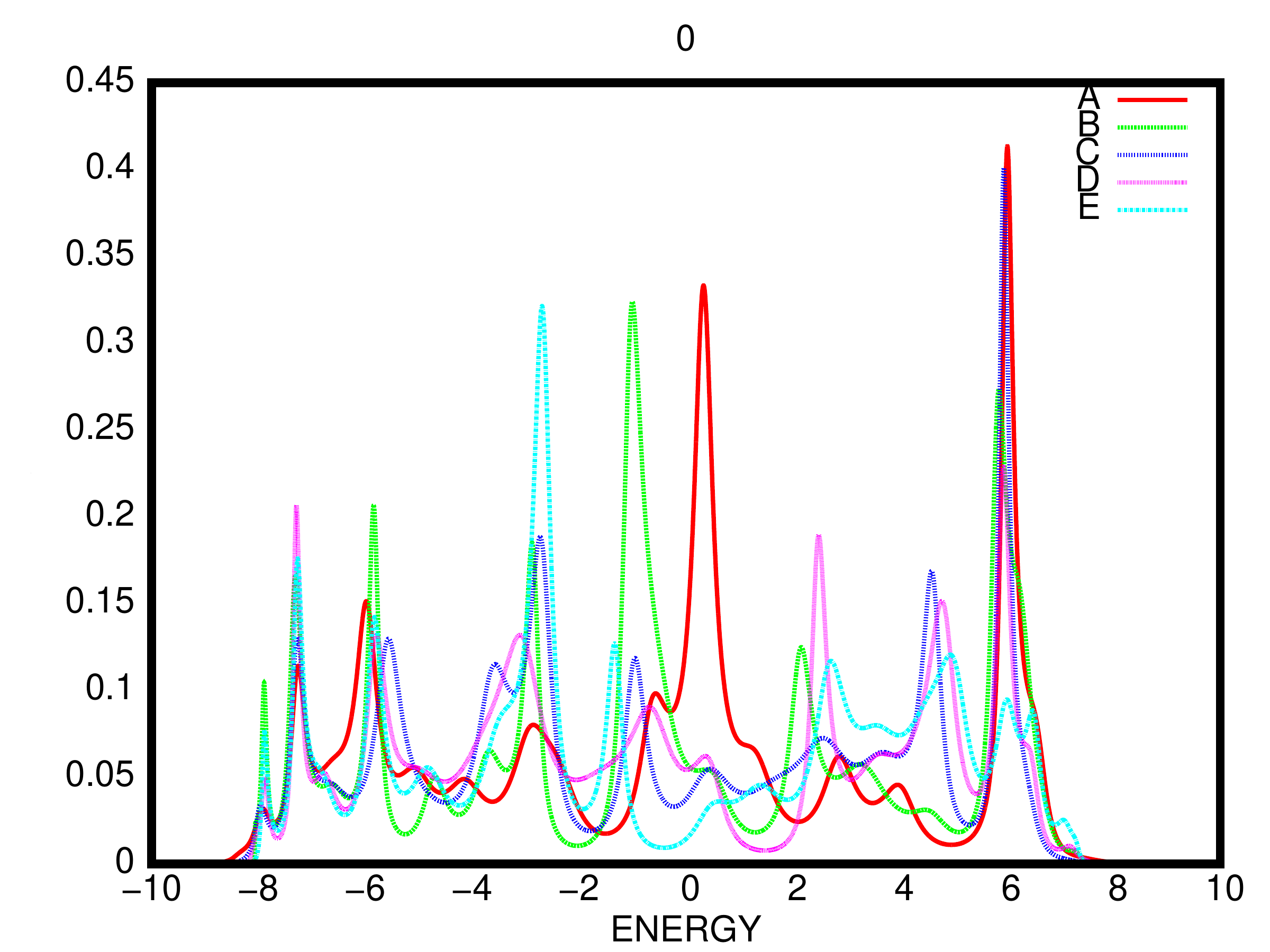}}
\subfigure[]{\includegraphics[width=3.7cm,height=4cm]{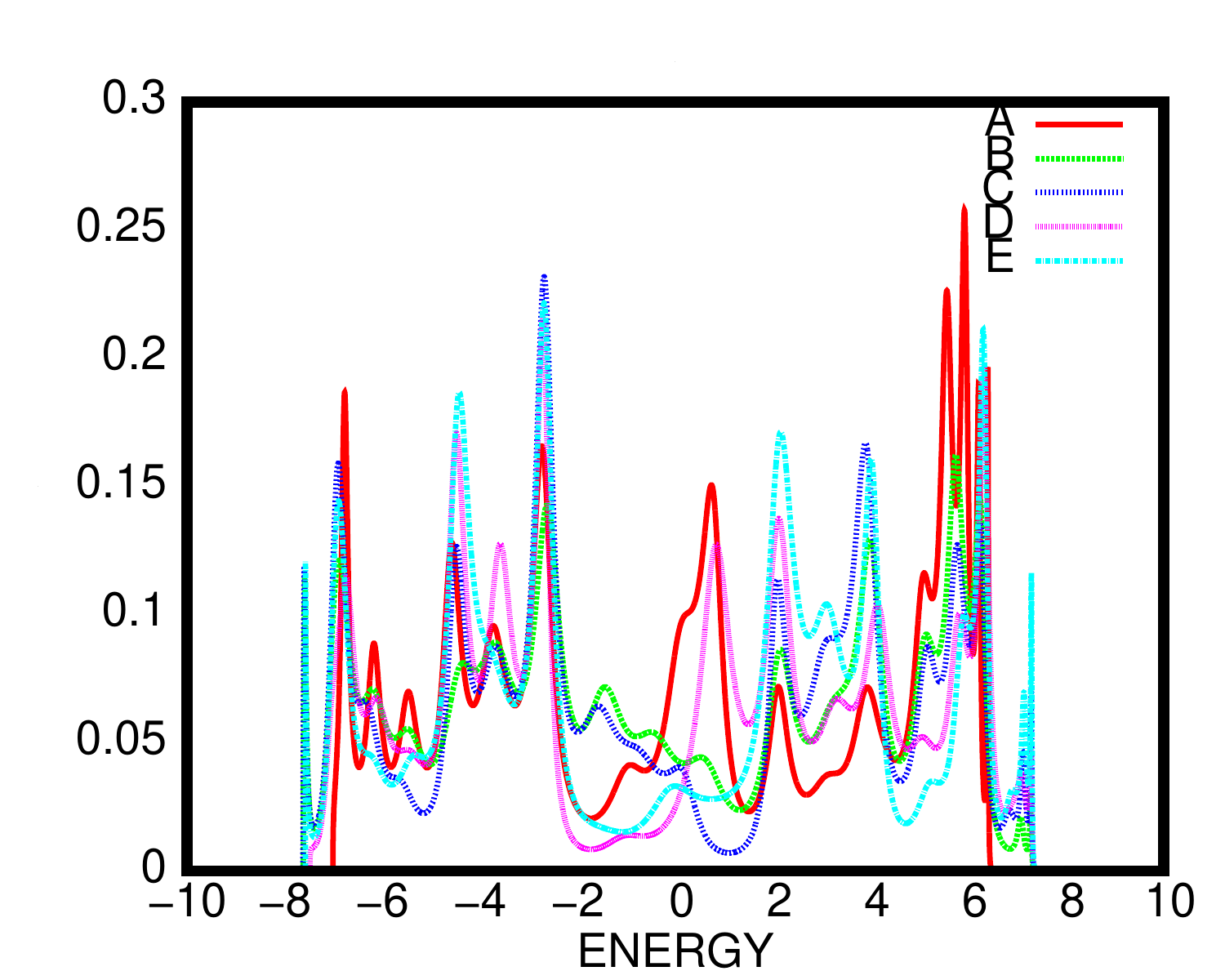}}
\subfigure[]{\includegraphics[width=3.7cm,height=4cm]{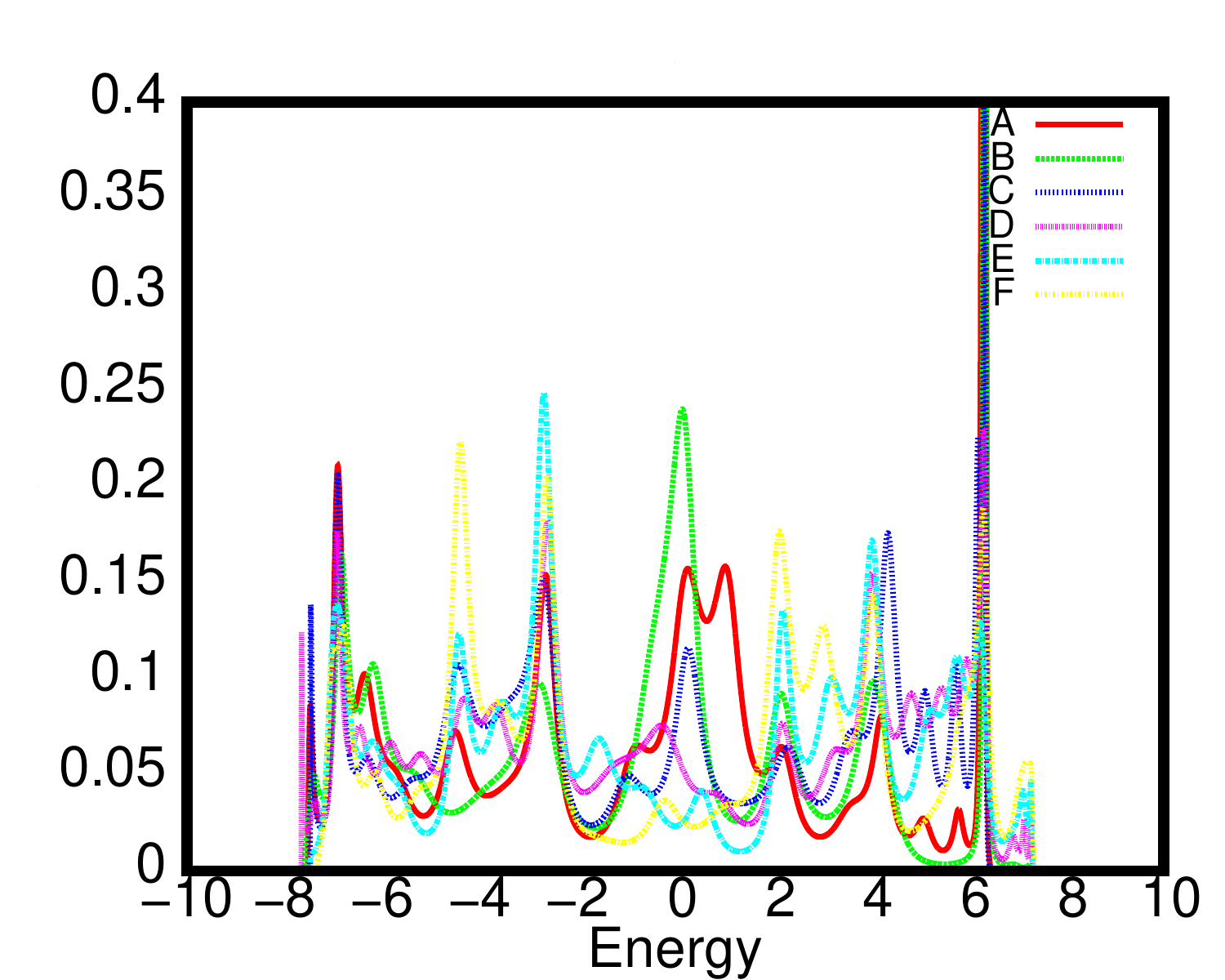}}
\caption{\small\label{gdos}p-z Projected LDOS in unit of /eV/atom for Graphene sheet with (a)$\epsilon=0,t=-0.2.544$(b)$\epsilon=-0.291,t=-2.544$,
LDOS for Single Layer Infinite Pristine Graphene Nanotubes of (c)(6,0) Zigzag type, (d)(3,3) Armchair type, (e)(6,6) Armchair type; Zigzag nanotube with (f)single S-W defect, (g)double S-W defects; Armchair nanotubes with (h)single S-W defect, (i)double S-W defects. (Colours: A-Red; B-Green; C-Blue; D-Pink; E-Cyan; F-Yellow.)}
\end{figure*}  
\pagebreak

 Fig.\ref{gdos}(a) gives the DOS of pristine graphene sheet with hopping parameter $t=-2.544$ and the potential energy term  $\epsilon=0$ (not the calculated value). This is symmetric about the minima, i.e., $E=0$ and the maxima are at $E= \pm t$. This DOS is similar to those calculated by other methods earlier. The only difference is that we found the peaks at $E=\pm 2.544 $ and in those papers, they are at $E =\pm 1$ as, in most of the papers the values that they put are $\epsilon= 0$ and $t=1$. Now, putting the calculated value of $\epsilon=-0.291$ we find Fig.\ref{gdos}(b). As expected, the entire graph shifts towards left by an amount $E= 0.291 $ and is symmetric about  $E=-0.291$.


\flushleft
\begin{table*}[htp]\label{ene}
\caption{\bf {Energies of a portion of Nanotubes using LMTO}}
\vskip 1cm
\begin{center}
\small
\begin{tabular}{|l|c|c|c|c|c|}
\hline
\centering
\bf {Nanotubes} & \bf {Radius(\rm \AA)} & \bf {No. of} & \bf {Total Energy} & \bf {Energy diff. wrt}  \\
 & & \bf { Atoms} & \bf {(Ry.)} & \bf {Graphene(Ry/atom)} \\ \hline
Planer Graphene & $\infty$ & 2 & -150.01542582 & 0.00 \\ \hline
Zigzag (6,0) & 4.919 & 72 & -5380.76910597 & 0.27480866 \\ \hline
Zigzag (6,0) + one SW defect & 4.919 & 72 & -5381.98834539 & 0.25787478 \\ \hline
Armchair (3,3) & 4.010 & 72 & -5390.14579977 & 0.14457680\\ \hline
Armchair (3,3) + one SW defect & 4.010 & 72 & -5390.13627404 & 0.14470910 \\ \hline
(6,0)+ one SW relaxed structure & N.A. & 72 & -5361.77814788 & 0.53857197 \\ \hline
\end{tabular}
\end{center}
\end{table*}

\begin{figure*}[b!]
 \centering
\subfigure[]{\framebox {\includegraphics[width=3cm,height=3cm]{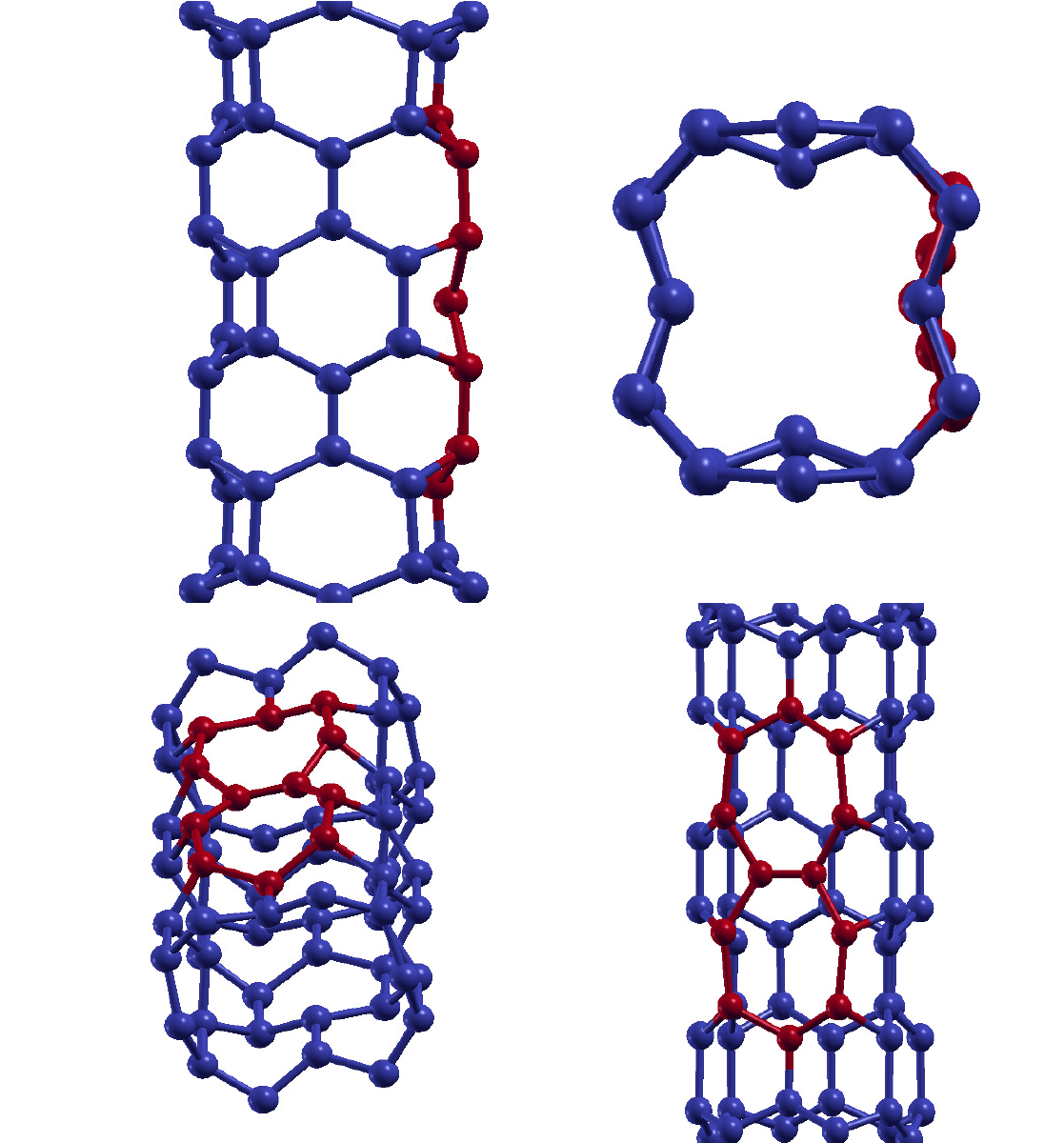}}}
\subfigure[]{\framebox {\includegraphics[width=3cm,height=3cm]{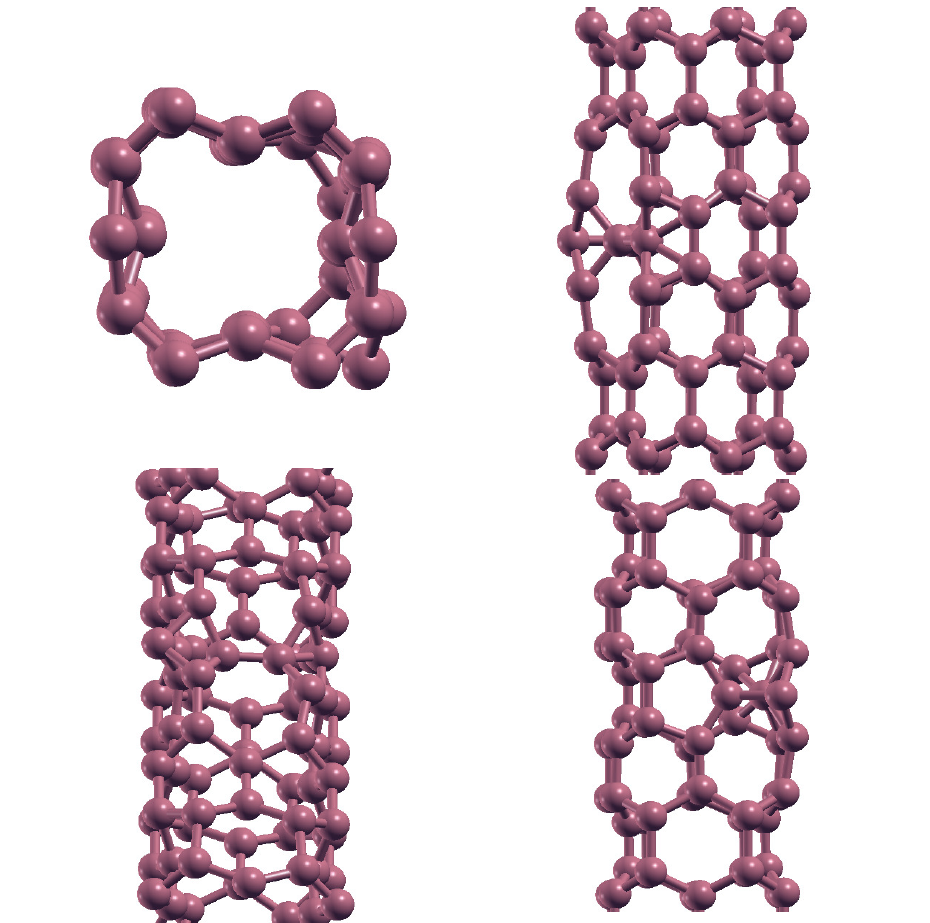}}}
\caption{\small\label{bomd}  Relaxed structure of zigzag nanotubes with (a)single S-W defect, (b)double S-W defects from different angles.}
\end{figure*}

\begin{figure*}[bh!]
\centering
\vskip -3cm
\subfigure[]{\framebox{\includegraphics[width=5.5cm,height=4cm]{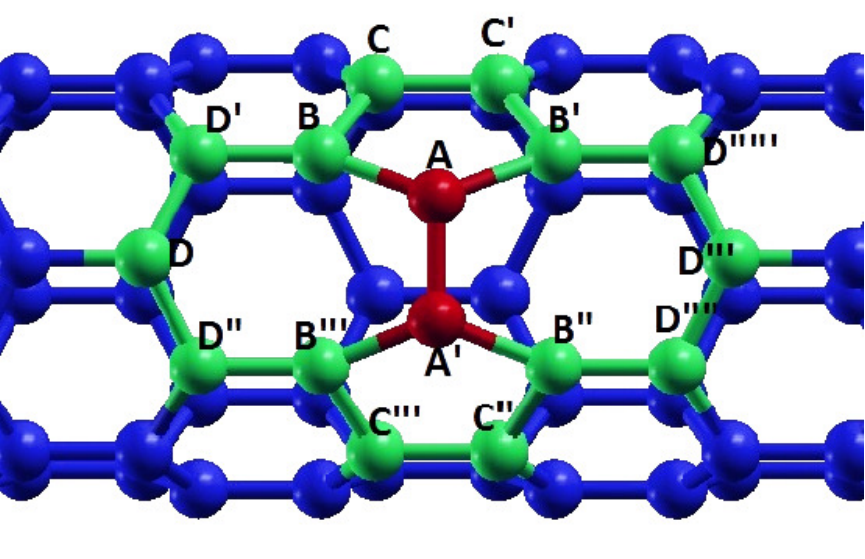}}}  
\subfigure[]{\framebox{\includegraphics[width=5.5cm,height=4cm]{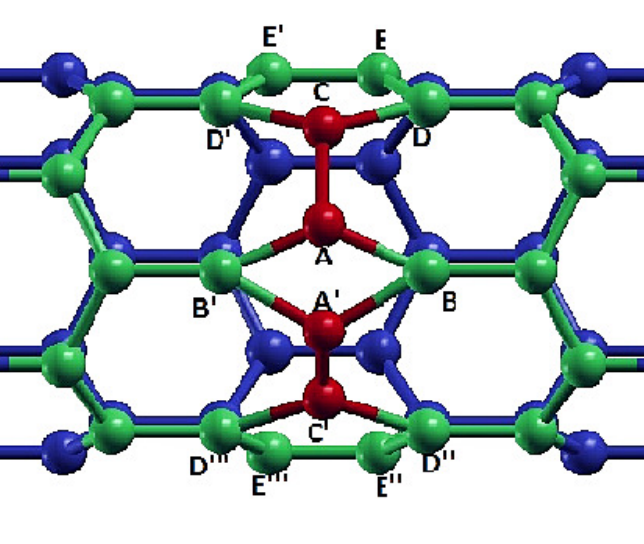}}}\\
\subfigure[]{\framebox{\includegraphics[width=5.5cm,height=3.5cm]{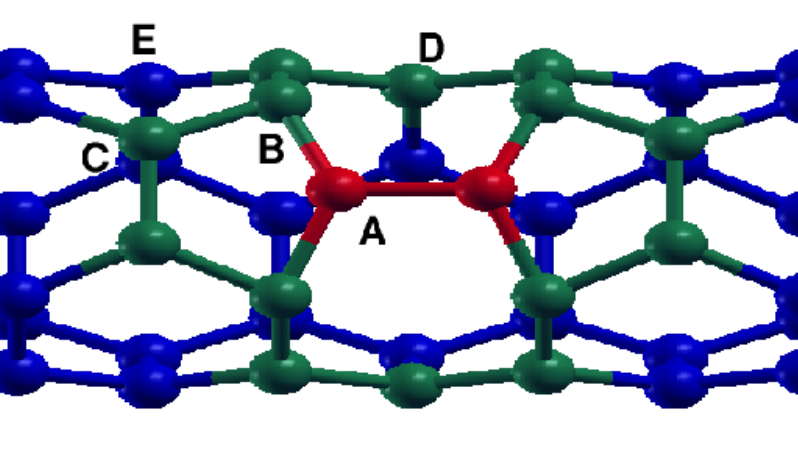}}}
\subfigure[]{\framebox{\includegraphics[width=5.5cm,height=3.5cm]{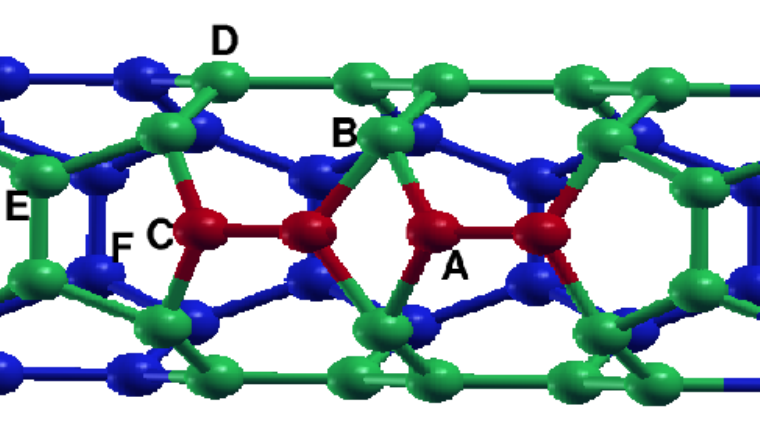}}}\\
\subfigure[]{\includegraphics[width=3.5cm,height=6.5cm]{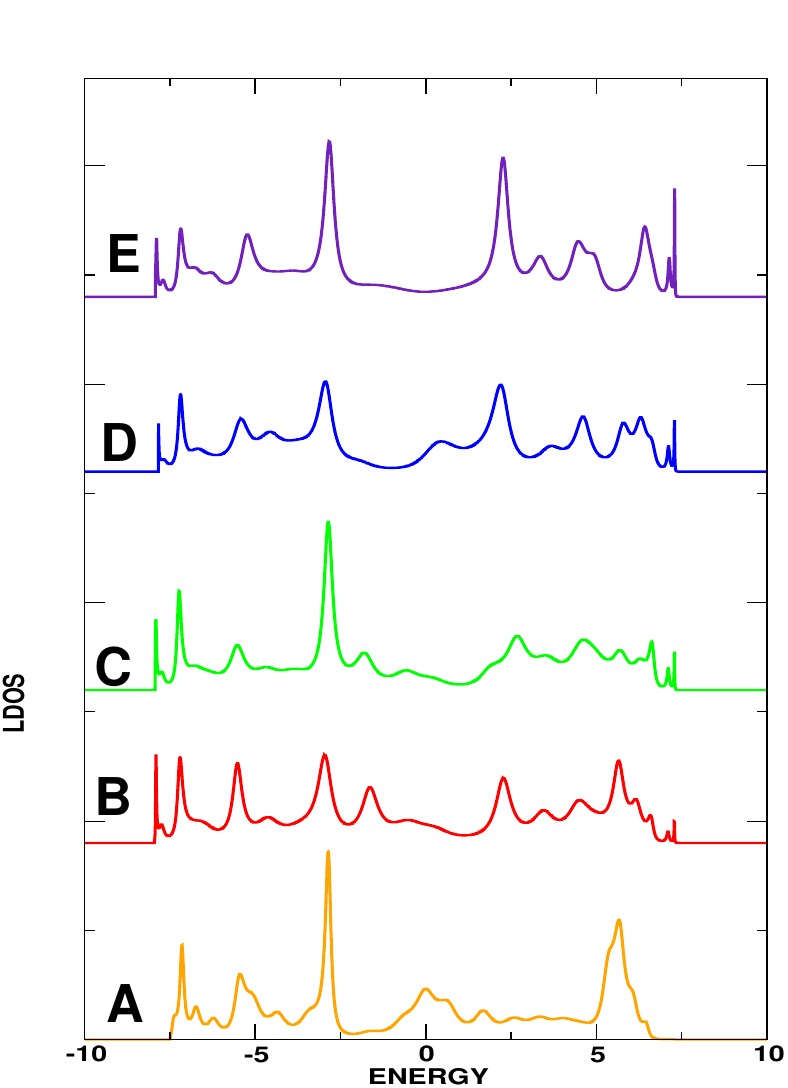}}    
\subfigure[]{\includegraphics[width=3.5cm,height=6.5cm]{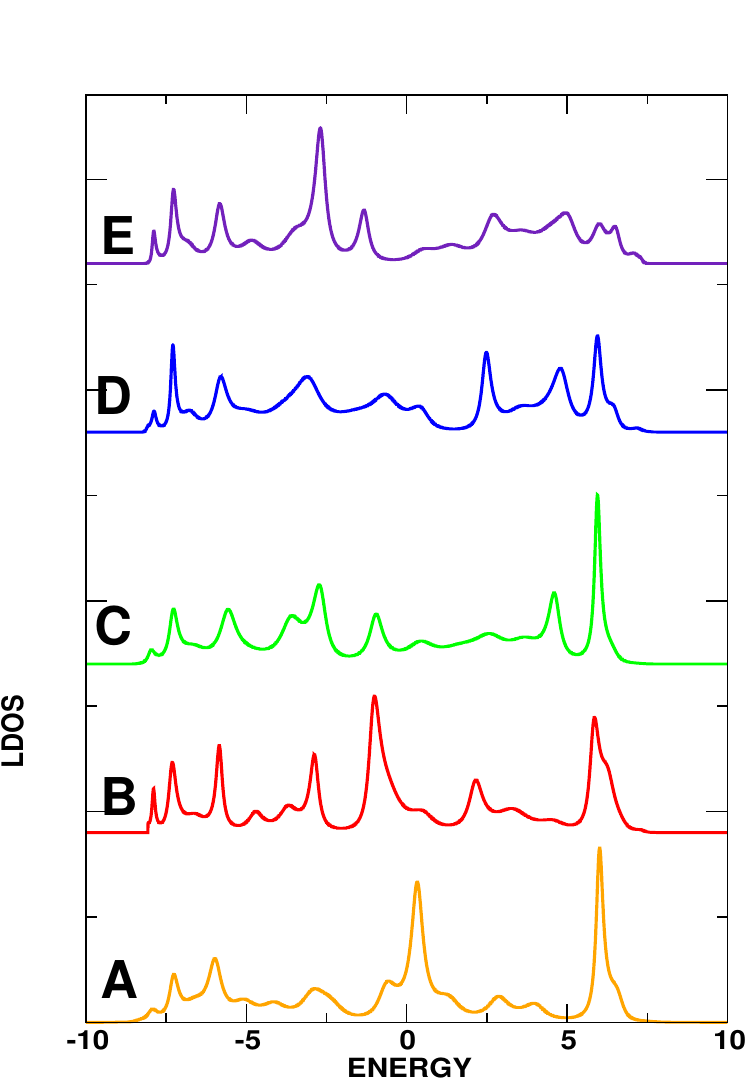}}
\subfigure[]{\includegraphics[width=3.5cm,height=6.5cm]{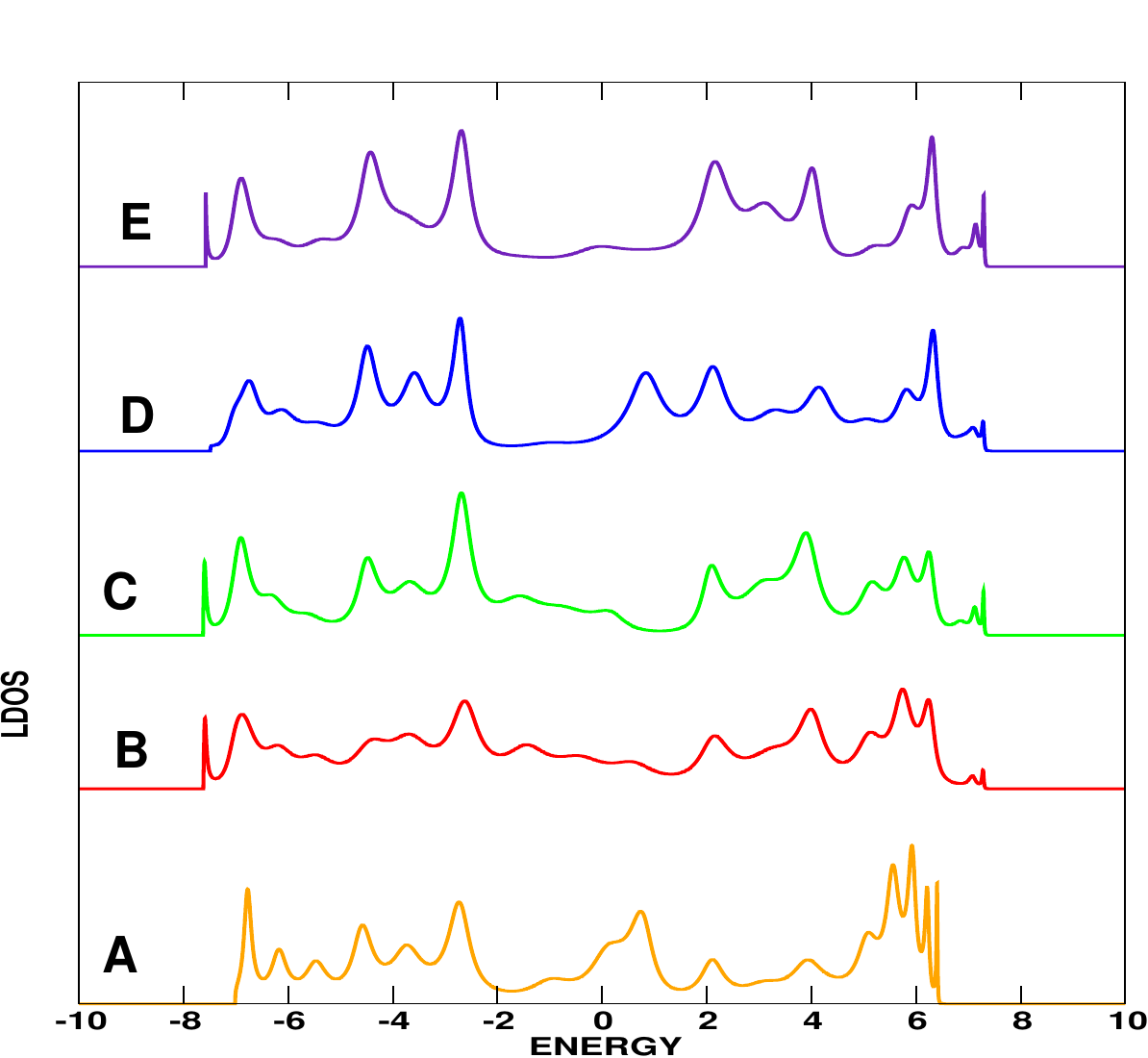}}    
\subfigure[]{\includegraphics[width=3.5cm,height=6.5cm]{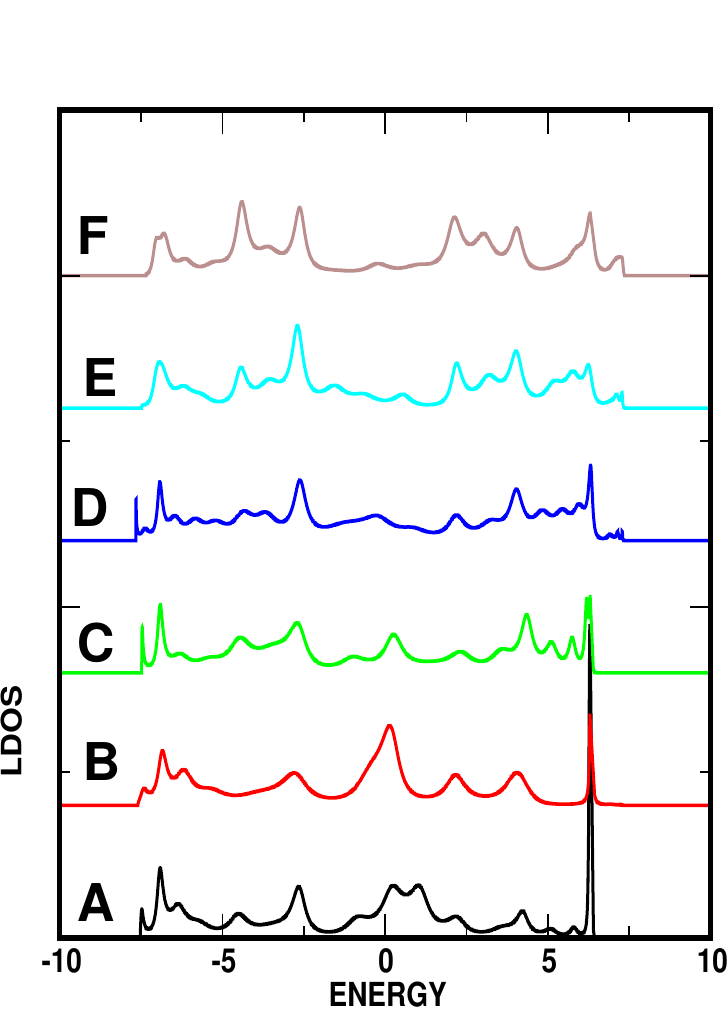}} 
\caption{\label{ldos} The positions at which the LDOS for (a)Zigzag nanotubes with single S-W defect,(b)Zigzag nanotubes with double S-W defects; (c)Armchair nanotubes with single S-W defect and (d)Armchair nanotubes with double S-W defects have been calculated; (e) to (h) the actual LDOS in unit of /eV/atom of these structures respectively.}
\end{figure*}

 A portion of the long nanotubes, each with 72 Carbon atoms, are cut and put into LMTO as the basis of a cell one by one. The calculated energies are given in Table 1. We see the S-W defect does not change the energy very much. This is understandable as, in the case of S-W defect, the bond lengths do not change appreciably. Now, these portion of the nanotubes with S-W defects are taken separately and first principle molecular dynamical relaxation are done. The resulted structures are given in Fig.\ref{bomd}(a) for nanotube with a single S-W defect and in Fig.\ref{bomd}(b) for nanotube with double S-W defects. In both of these, the cylindrical structure is distorted. The calculated total energy of the structure in Fig.\ref{bomd}(a) is -5361.778 Ry., which is higher than the energy of the structure in Fig.\ref{grph}(c) by 20 Ry., approximately. This means, the structure in Fig.\ref{grph}(c) is more stable, all the further calculations are done on this structure. These calculated energies are for small samples, so, do not clearly represent the nanotube structures we started with. Our samples are much longer in length. Though LMTO and NMTO can not handle any sample of such big size, they can calculate real space Hamiltonian parameters for pristine structures. These parameters are used in the recursion method and can produce good results. The LDOS for different atoms of the nanotubes are given in Fig.\ref{gdos} (f) to (i) and replotted in Fig.\ref{ldos} for better understanding.  
 
 In Fig.\ref{gdos} the LDOS for pristine nanotube of (c) (6,0) zigzag type, (d) (3,3) armchair types and (e) (6,6) armchair types are shown. There are some geometric distinctions between the infinite flat sheet and the nanotube. Both geometries have no open edges, though, in the tube, this is achieved by periodicity along one edge and shifting another edge to infinity; while for the sheet, all edges are shifted to infinity. In spite of these similarities, the LDOS are quite different as seen in Fig.\ref{gdos}. As we compare Fig.\ref{gdos}(b) with Fig.\ref{gdos}(d) and (e), we find LDOS of graphene sheet has more differences with LDOS of (3,3) armchair nanotube as compared to the  (6,6) type. This is expected as nanotubes with higher chirality has surface with less curvature, so, has fewer dissimilarities with the infinite sheet.

  We can understand this from the recursion coefficients. In pristine graphene $\alpha_n$ are independent of $n$ and consequently all odd moments of Green's function are zero. This leads to LDOS symmetric about the origin. However, to interpret Fig \ref{gdos}(d) and (e) correctly, we compare Fig \ref{ab}(a) with (b) and (c); and note that in the latter  $\alpha_n$ oscillates with $n$ while in the former it is a constant. This leads to asymmetric structures in the LDOS for graphene with defects. When topological defects are introduced, the geometrical symmetry is broken, so, the diagonal term of TB-Hamiltonian plays a big role in determining the exact shape of the LDOS. We should like to emphasize this point so that the general practice of shifting the origin to a $\epsilon=0$ is discouraged.

 LDOS at the point E of zigzag type nanotube with a single S-W defect (refer to Fig.\ref{ldos}(e)) closely resembles Fig.\ref{gdos}(c) with a semiconductor like feature near Fermi energy ($E_F$). As we enter the defect region, this energy range of low LDOS is filled up with impurity states. This is also true for nanotubes with double S-W defects (refer to Fig.\ref{ldos}(f)). In the case of the armchair nanotube, LDOS at E is more metallic in nature (Fig.\ref{ldos}(g)). As we get near to the defect, LDOS at $E_F$ increases, leading to greater metallicity. The effect of the double S-W defects (Fig.\ref{ldos}(h)) is rather similar except that the metallicity increases further. From this, we can surmise that incorporation of S-W defects should increase the metallicity in Carbon nanotubes. However, we are not in a position yet of generalising the statement, as, in the case of planer graphene sheets with S-W defects, no such prominent metallic behaviours was observed \cite{am1}.

\section{Conclusion}
 
 We have shown that the recursion method is ideally suited for the study of electronic structure in graphinic materials with extended defects. This method coupled with the TB-LMTO/NMTO provides a first principles description, and our application to graphene nanotubes justifies this view. Moreover, the same recursion method, suitably modified, can be used to study response functions like resistivity and optical conductivity \cite{opt}, phonons and thermal conduction  \cite{alam} and magnetism. We propose this unified technique for our future work. 

\section{Acknowledgement}
 The authors would like to thank Dhani Nafday and Tanusri Saha-Dasgupta for providing us with the Hamiltonian parameters calculated by NMTO. Authors also like to thank Mukul Kabir for fruitful discussions.

\pagebreak 
\pagebreak
\newpage

{\baselineskip 10pt
\small\rm

}
 \end{document}